\documentclass[lettersize,journal]{IEEEtran}
\usepackage{amsmath,amsfonts}
\usepackage{algorithmic}
\usepackage{algorithm}
\usepackage{array}
\usepackage{textcomp}
\usepackage{stfloats}
\usepackage{url}
\usepackage{verbatim}
\usepackage{graphicx}
\usepackage{indentfirst}
\usepackage{multirow}
\usepackage{graphicx}
\usepackage{color}
\usepackage{xcolor}
\usepackage{colortbl}
\usepackage{listings}
\usepackage{enumerate}
\usepackage{amsmath,amsthm,amscd,amssymb,latexsym,upref,stmaryrd}
\usepackage{amsfonts,epsfig}
\usepackage{CJK}
\usepackage{float}
\usepackage{lipsum}
\usepackage{multicol}
\usepackage{setspace}
\usepackage{bm}
\usepackage{amsthm}
\usepackage{amsmath}
\usepackage{amssymb}
\usepackage{amsfonts}
\usepackage{arydshln}
\usepackage{stfloats}
\usepackage{booktabs}
\usepackage{threeparttable}
\usepackage{stackengine}
\usepackage[utf8]{inputenc}
\usepackage{fancyhdr}
\usepackage[colorlinks,linkcolor=black,anchorcolor=black,citecolor=black,urlcolor=black]{hyperref}

\usepackage{makecell}
\usepackage{subcaption}

\newcommand{\xrowht}[2][0]{\addstackgap[.5\dimexpr#2\relax]{\vphantom{#1}}}

\newtheorem{remark}{Remark}

\begin{document}

\title{Integrated Communication and Learned Recognizer with Customized RIS Phases and Sensing Durations}

\author{Yixuan~Huang,~\IEEEmembership{Graduate Student Member,~IEEE,} Jie~Yang,~\IEEEmembership{Member,~IEEE,} Chao-Kai~Wen,~\IEEEmembership{Fellow,~IEEE,} and Shi~Jin,~\IEEEmembership{Fellow,~IEEE}
\thanks{
Manuscript received 31 May 2024; revised 28 October 2024 and 16 January 2025; accepted 28 Febuary 2025.
Date of publication ** March 2025; date of current version ** **** 2025.
This work was supported
in part by the National Natural Science Foundation of China (NSFC) under Grant 62261160576, Grant 624B2036, and Grant 62301156;
in part by the Fundamental Research Funds for the Central Universities under Grant 2242022k60004 and Grant 2242023K5003;
in part by the Key Technologies R\&D Program of Jiangsu (Prospective and Key Technologies for Industry) under Grant BE2023022-1 and Grant BE2023022;
in part by the Postgraduate Research and Practice Innovation Program of Jiangsu Province under Grant KYCX24\_0402;
in part by the Southeast University Innovation Capability Enhancement Plan for Doctoral Students under Grant CXJH\_SEU 24082 and Grant CXJH\_SEU 25018.
The work of C.-K. Wen was supported in part by the National Science and Technology Council of Taiwan through grant MOST 111-2221-E-110-020-MY3, and by the Sixth Generation Communication and Sensing Research Center, which is funded by the Higher Education SPROUT Project of the Ministry of Education of Taiwan.
Part of this paper has been presented at the International Conference on Future Communications and Networks (FCN), St Julian's, Malta, November 2024, and received the Best Student Paper Award \cite{huang2024learned}.
The associate editor coordinating the review of this article and approving it for publication was Y. Shen. (Corresponding authors: Jie Yang; Shi Jin.)

Yixuan Huang is with the National Mobile Communications Research Laboratory, Southeast University, Nanjing 210096, China (e-mail: huangyx@seu.edu.cn).

Jie Yang is with the Key Laboratory of Measurement and Control of Complex Systems of Engineering, Ministry of Education, and the Frontiers Science Center for Mobile Information Communication and Security, Southeast University, Nanjing 210096, China (e-mail: yangjie@seu.edu.cn).

Chao-Kai Wen is with the Institute of Communications Engineering, National Sun Yat-sen University, Kaohsiung 80424, Taiwan (e-mail: chaokai.wen@mail.nsysu.edu.tw).

Shi Jin is with the National Mobile Communications Research Laboratory, and the Frontiers Science Center for Mobile Information Communication and Security, Southeast University, Nanjing 210096, China (e-mail: jinshi@seu.edu.cn).

Color versions of one or more figures in this article are available at https://doi.org/10.1109/TCOMM***

Digital Object Identifier 10.1109/TCOMM***}
}

\maketitle

\begin{abstract}

Future wireless communication networks are expected to be smarter and more aware of their surroundings, enabling a wide range of context-aware applications. Reconfigurable intelligent surfaces (RISs) are set to play a critical role in supporting various sensing tasks, such as target recognition. However, current methods typically use RIS configurations optimized once and applied over fixed sensing durations, limiting their ability to adapt to different targets and reducing sensing accuracy.
To overcome these limitations, this study proposes an advanced wireless communication system that multiplexes downlink signals for environmental sensing and introduces an intelligent recognizer powered by deep learning techniques.
Specifically, we design a novel neural network based on the long short-term memory architecture and the physical channel model. This network iteratively captures and fuses information from previous measurements, adaptively customizing RIS phases to gather the most relevant information for the recognition task at subsequent moments. These configurations are dynamically adjusted according to scene, task, target, and quantization priors.
Furthermore, the recognizer includes a decision-making module that dynamically allocates different sensing durations, determining whether to continue or terminate the sensing process based on the collected measurements. This approach maximizes resource utilization efficiency. Simulation results demonstrate that the proposed method significantly outperforms state-of-the-art techniques while minimizing the impact on communication performance, even when sensing and communication occur simultaneously.
Part of the source code for this paper can be accessed at \url{https://github.com/kiwi1944/CRISense}.

\end{abstract}


\section{Introduction}
Environment sensing is expected to be integrated into future wireless communication systems to enable ubiquitous sensing using channel state information (CSI), facilitating tasks such as mapping, imaging, and recognition \cite{liu2024senscap,win2022location,kim2023ris}. Among these, target recognition has emerged as a critical issue for supporting context-aware applications. For example, health monitoring and touchless human-computer interaction are made possible through human posture recognition \cite{hu2020reconfigurable}, while classifying birds and drones enhances security surveillance \cite{duan2023classification}. This study primarily investigates advanced classification techniques, without restricting them to specific applications.

Classification is a traditional issue that has been extensively explored in the computer vision (CV) field \cite{mnih2014recurrent}, inspiring some prior studies to design radio classifiers by first imaging the targets and then classifying their radio images \cite{saigre2022intelligent,kim2021human}. However, radio imaging is inherently challenging, typically requires specific sensing systems, and leads to high measurement acquisition and digital processing burdens \cite{zhu2016frequency,huang2024fourier}. Moreover, extensive CSI measurements are acquired to capture detailed information about the target images, of which only a small portion is relevant to the recognition task \cite{saigre2022intelligent}. Thus, designing classifiers that directly map the limited measurements to class labels without imaging is considered more efficient. Human posture recognition systems have been proposed based on radio frequency identification transceivers by classifying the received signal strength \cite{yao2015rf}. Multiple-input multiple-output radars are employed for target recognition based on radar cross-section (RCS) estimates \cite{sasakawa2018human}. Since communication systems ubiquitously transmit and receive radio signals, CSI can be acquired during regular communication. In this study, we leverage a communication system with a full-duplex base station (BS) that multiplexes the echo of downlink (DL) signals to derive CSI for target recognition \cite{lu2024random,mehrotra2022degrees}.

Despite these advancements, the complex and unpredictable nature of wireless channels inherently limits sensing accuracy. Recently, reconfigurable intelligent surfaces (RISs) have been used to tailor electromagnetic environments for communication and sensing with low hardware costs and energy consumption \cite{guo2022ris,wu2023design,qian2023optimization,wang2022location}. In RIS-aided systems, the designs of the classification function and RIS phases have been two critical issues. Concerning the classification function design, a hypothesis-testing-based method has been proposed in \cite{dos2024assessing}, which incurs high complexity when calculating posterior probabilities across a large number of categories. In contrast, deep learning-based techniques employing fully connected (FC) neural networks (NNs) have been utilized in \cite{hu2020reconfigurable} to effectively fit the classification function, reducing complexity and significantly enhancing accuracy. Thus, NN-based techniques are employed for target recognition in this study.

For RIS phase design, random configurations are utilized to gather diverse information about the target during sensing \cite{zhao2023intelligent}. However, these configurations indiscriminately collect all available information, ignoring useful prior knowledge that could improve recognition performance. Analogous to radio imaging systems that minimize the average mutual coherence of the sensing matrix \cite{huang2024ris}, similar RIS phase designs have been employed in \cite{hu2020reconfigurable}. Alternatively, a principal component analysis-based dictionary can be trained to optimize RIS radiation patterns to illuminate the principal components of a given scene \cite{li2019machine}. While these methods \cite{hu2020reconfigurable,li2019machine} incorporate prior knowledge about the scene in RIS phase design, they ignore task-specific information and do not optimize RIS phases directly for target recognition. Furthermore, these studies optimize measurement acquisition and processing independently, neglecting the close relationship between them.

\begin{table*}[t]
  \renewcommand{\arraystretch}{1.4}
  \centering
  \fontsize{8}{8}\selectfont
  \captionsetup{font=small}
  \caption{Comparison with prior studies.}\label{tab-compare}
  \setlength{\arrayrulewidth}{0.4mm}
  \begin{threeparttable}
    \begin{tabular}{|c|p{1.3cm}<{\centering}|p{1.3cm}<{\centering}|p{1.3cm}<{\centering}|p{1.3cm}<{\centering}|c|c|}
      \hline
      \multirow{2}{*}{Methods} & \multicolumn{4}{c|}{Prior information utilization} & \multirow{2}{*}{\makecell[c]{RIS phase \\ generation mode}} & \multirow{2}{*}{\makecell[c]{Support for dynamic \\ sensing durations}} \\
      \cline{2-5}
       & Scene & Task & Quantization & Target & & \\\cline{2-5}
      \hline
      \cite{zhao2023intelligent} & & & & & Random & \\
      \hline
      \cite{hu2020reconfigurable,li2019machine} & \checkmark & & & & Disposably optimized & \\
      \hline
      \cite{hu2021metasensing,del2020learned} & \checkmark & \checkmark & \checkmark & & Disposably optimized & \\
      \hline
      Proposed & \checkmark & \checkmark & \checkmark & \checkmark & Dynamically generated & \checkmark \\
      \hline
    \end{tabular}
  \end{threeparttable}
\end{table*}

To address these challenges, a method that jointly optimizes RIS phases and NN parameters is studied using reinforcement learning (RL) methods \cite{hu2021metasensing}. Additionally, a learned integrated sensing pipeline (LISP) is proposed in \cite{del2020learned}, integrating RIS phases as trainable physical variables within the NN. As a result, RIS phases and NN parameters are jointly optimized through supervised learning, achieving state-of-the-art target recognition performance. However, prior methods \cite{hu2020reconfigurable,li2019machine,hu2021metasensing,del2020learned} optimize all RIS configurations simultaneously, without considering that prior measurements contain useful information about the target. 
In \cite{sohrabi2022active}, historical CSI measurements are employed to assist in designing RIS phases in subsequent intervals to improve communication performance, which is similar to the concept of the recurrent visual attention model in the CV field \cite{mnih2014recurrent}. Inspired by \cite{sohrabi2022active,mnih2014recurrent}, a long short-term memory (LSTM) network is employed to customize RIS phases for precise target recognition by leveraging information from previous measurements. The optimized RIS phases are implicitly embedded into NN parameters and derived through NN inference.

In recognition-based applications, minimizing the required number of measurements while ensuring sensing accuracy can improve multiple metrics, including enhanced sensing speed and reduced communication resource occupation \cite{huang2024fourier,del2020learned}.
Since different targets exhibit unique scattering properties, the difficulty of classification may vary. Fewer measurements may suffice for ``easy'' targets, while ``hard'' targets may require more measurements for reliable classification. However, this property has not been exploited in previous studies, which use fixed numbers of RIS configurations \cite{hu2020reconfigurable,zhao2023intelligent,li2019machine,hu2021metasensing,del2020learned}. Additionally, FC networks, which process all CSI measurements simultaneously, are incompatible with dynamic sensing durations due to the changing number of input neurons.

Customizing inference durations is not a new concept in deep learning. In FC networks and convolutional NNs (CNNs), early exiting mechanisms have been added to hidden layers to control the continuation or termination of the inference process \cite{liu2023resource,xu2023lgvit}, enabling dynamic inference times. Using a recurrent NN (RNN), \cite{wang2020glance} proposed terminating the inference process when the classifier's maximum probability exceeds a certain threshold. Additionally, RL techniques can be used to train a decision layer that automatically decides when to stop inference \cite{li2017dynamic}.
Inspired by these deep learning innovations, we propose to dynamically customize the sensing duration for each target to minimize the required number of measurements. Specifically, we enhance the proposed NN by introducing a decision module that interacts with the LSTM network, analyzing acquired information to determine whether to continue or stop sensing.

The comparison with prior studies is summarized in Table \ref{tab-compare}. Our key contributions are as follows:

\begin{itemize}
\item \textbf{Design of an integrated communication and recognition protocol:} We propose a novel RIS-aided communication system that embeds recognition functionality by multiplexing DL signals. The protocol designed for integrated communication and recognition optimizes RIS phases for communication and sensing in time-division modes, achieving high sensing accuracy with minimal impact on communication performance.

\item \textbf{Joint optimization of NN parameters and adaptively customized RIS phases:} We introduce an innovative NN based on LSTM and the physical channel model, which adaptively customizes RIS phase shifts at each moment by leveraging prior information from acquired measurements. RIS phases are unique to each target, and NN parameters and RIS phases are jointly optimized using scene, task, target, and quantization priors through supervised learning, significantly improving recognition accuracy compared to the state-of-the-art method \cite{del2020learned}. 

\item \textbf{Proposal of dynamic sensing durations:} We further introduce a decision module to the proposed NN that automatically terminates the sensing process, allocating distinct measurement counts to different targets. The NN is trained based on RL inspirations and a curriculum learning strategy. The decision module improves sensing accuracy while reducing the required number of measurements, speeding up the sensing process and minimizing communication resource usage.
\end{itemize}

{\bf Notations}---The scalars (e.g., $a$) are denoted in italics, vectors (e.g., $\mathbf{a}$) in bold, and matrices (e.g., $\mathbf{A}$) in bold capital letters. The $\ell_{2}$-norm of $\mathbf{a}$ is given as $\|\mathbf{a}\|_{2}$. $|a|$ takes the modulus of $a$, and $j = \sqrt{-1}$ is the imaginary unit. $\text{diag}{(\mathbf{a})}$ constructs a diagonal matrix with the elements of $\mathbf{a}$. Transpose and Hermitian operators are $(\cdot)^{\text{T}}$ and $(\cdot)^{\text{H}}$, respectively.

\section{System Model}
\label{sec:system-model}

\begin{figure}
    \centering
    \includegraphics[width=0.75\linewidth]{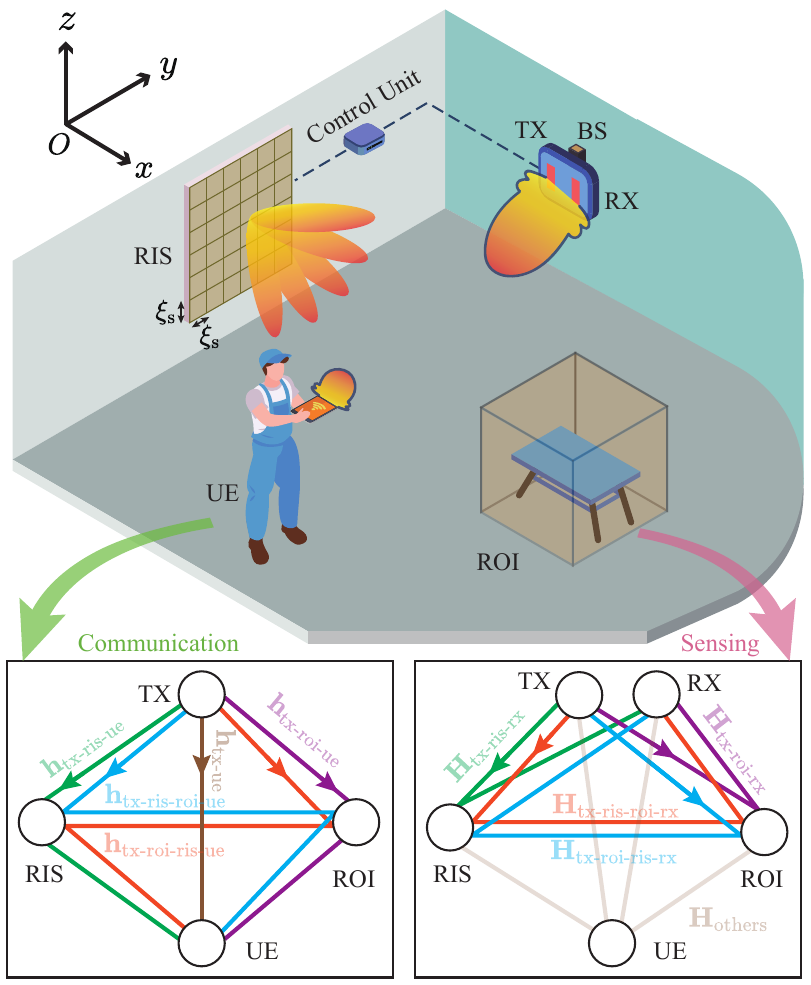}
    \captionsetup{font=footnotesize}
     \caption{Illustration of the proposed RIS-aided joint communication and recognition system.} 
    \label{fig-model}
\end{figure}

We consider a RIS-aided communication system functioning within the 3D space $[x,y,z]^{\text{T}} \in \mathbb{R}^3$, as illustrated in Fig. \ref{fig-model}. The full-duplex BS communicates with a single-antenna user equipment (UE) using orthogonal frequency division multiplexing (OFDM) signals.
The system employs $N_{\text{f}}$ subcarriers, spaced by $\Delta f$.
Although multi-user communication can be supported via orthogonal frequency division multiple access technology, for simplicity, we focus on a single UE. The BS's transmitter (TX) and receiver (RX) consist of uniform linear arrays with $N_{\text{t}}$ and $N_{\text{r}}$ antennas, respectively, both spaced at $\lambda_0/2$, where $\lambda_0$ is the wavelength of the center subcarrier at frequency $f_0$.
The RIS consists of $N_{\text{s}}$ elements, each of size $\xi_{\text{s}}\times\xi_{\text{s}}$. Its phase shifts, represented as $\boldsymbol{\omega} = [\omega_1, \omega_2, \ldots, \omega_{N_{\text{s}}}]^{\text{T}} \in \mathbb{C}^{N_{\text{s}} \times 1}$, are adjusted by the BS in real time \cite{li2020intelligent}.
The large RIS array can capture rich target information, enabling high sensing performance.

The positions of the TX, RX, and RIS are assumed to be known.
The region of interest (ROI) is a large area where potential targets may be located. The center of the ROI is predefined at a distance $D$ from the RIS \cite{hu2020reconfigurable,wang2024dreamer}.
The ROI can be discretized into $N_{\text{i}}$ voxels, and its radio image is represented by $\boldsymbol{\sigma} = [\sigma_1, \sigma_2, \ldots, \sigma_{N_{\text{i}}}]^{\text{T}} \in \mathbb{R}^{N_{\text{i}} \times 1}$, where $\sigma_{n_{\text{i}}}$ denotes the scattering coefficient of the $n_{\text{i}}$-th voxel, and $\sigma_{n_{\text{i}}} = 0$ if the $n_{\text{i}}$-th voxel contains no objects. 
Since the communication bandwidth is typically much smaller than the carrier frequency, we assume that the RIS phase shifts $\boldsymbol{\omega}$ and the ROI image $\boldsymbol{\sigma}$ remain constant across the $N_{\text{f}}$ subcarriers \cite{huang2024fourier,wu2023design}.
The goal of this study is to identify the class label of the target within the ROI during the communication process.

\subsection{Signal and Channel Models for Communication}

For communication, we consider a DL scenario, where the TX sends the signal $\mathbf{x}_{n_{\text{f}}} \in \mathbb{C}^{N_{\text{t}} \times 1}$ to the UE on the $n_{\text{f}}$-th subcarrier.
The corresponding received signal at the UE is expressed as
\begin{equation}\label{eq-commun-signal}
{r}_{\text{com}, n_{\text{f}}} = \sqrt{P_{\text{t}, n_{\text{f}}}}\,\mathbf{h}_{\text{com}, n_{\text{f}}}^{\text{T}}\mathbf{x}_{n_{\text{f}}} + {z}_{\text{com}, n_{\text{f}}},
\end{equation}
\hspace{-0.25em}where $\mathbf{h}_{\text{com}, n_{\text{f}}}^{\text{T}}\in\mathbb{C}^{1\times N_{\text{t}}}$ denotes the multipath channel from the TX to the UE, and ${z}_{\text{com}, n_{\text{f}}}\in\mathbb{C}$ represents the additive Gaussian noise at the UE.
The transmit power allocated to the $n_{\text{f}}$-th subcarrier is denoted as $P_{\text{t}, n_{\text{f}}}$, and the total transmit power is $P_{\text{t}}=\sum_{n_{\text{f}}=1}^{N_{\text{f}}}P_{\text{t}, n_{\text{f}}}$.
For simplicity, we assume $P_{\text{t}, n_{\text{f}}}=P_{0}$, where $n_{\text{f}}=1,2,\ldots,N_{\text{f}}$, resulting in $P_{\text{t}}=N_{\text{f}}P_{0}$.
Additionally, we assume $\|\mathbf{x}_{n_{\text{f}}}\|_2 = 1$.
According to Fig. \ref{fig-model}, the channel $\mathbf{h}_{\text{com}, n_{\text{f}}}$ can be formulated as
\begin{equation}\label{eq-commun-channel}
\begin{aligned}
\mathbf{h}_{\text{com}, n_{\text{f}}} =  & \ \mathbf{h}_{\text{tx}\text{-}\text{ue}, n_{\text{f}}} + \mathbf{h}_{\text{tx}\text{-}\text{roi}\text{-}\text{ue}, n_{\text{f}}} + \mathbf{h}_{\text{tx}\text{-}\text{ris}\text{-}\text{ue}, n_{\text{f}}}\\
& + \mathbf{h}_{\text{tx}\text{-}\text{roi}\text{-}\text{ris}\text{-}\text{ue}, n_{\text{f}}} + \mathbf{h}_{\text{tx}\text{-}\text{ris}\text{-}\text{roi}\text{-}\text{ue}, n_{\text{f}}} + \mathbf{h}_{\text{others}, n_{\text{f}}},
\end{aligned}
\end{equation}
where $\mathbf{h}_{\text{tx}\text{-}\text{ue}, n_{\text{f}}}^{\text{T}}\in\mathbb{C}^{1\times N_{\text{t}}}$ denotes the line-of-sight (LOS) path from the TX to the UE. $\mathbf{h}_{\text{tx}\text{-}\text{ris}\text{-}\text{ue}, n_{\text{f}}}^{\text{T}}$ and $\mathbf{h}_{\text{tx}\text{-}\text{roi}\text{-}\text{ue}, n_{\text{f}}}^{\text{T}}$ represent the single-bounce paths scattered by the RIS and the target in the ROI, respectively. The terms $\mathbf{h}_{\text{tx}\text{-}\text{ris}\text{-}\text{roi}\text{-}\text{ue}, n_{\text{f}}}^{\text{T}}$ and $\mathbf{h}_{\text{tx}\text{-}\text{roi}\text{-}\text{ris}\text{-}\text{ue}, n_{\text{f}}}^{\text{T}}$ represent two-bounce paths. Detailed expressions for these cascaded channels are provided in Appendix \ref{appendix-channel}.
Multipath components involving more than two bounces and random scatterers are included in $\mathbf{h}_{\text{others}, n_{\text{f}}}$.

\subsection{Signal and Channel Models for Sensing}

For environmental sensing, we consider that the BS's transmit signal $\mathbf{x}_{n_{\text{f}}}$ can be simultaneously received by its RX.
After the scattering of the RIS and the target, the received signals at the RX are given by 
\begin{equation}
\mathbf{r}_{\text{sen}, n_{\text{f}}} = \sqrt{P_{\text{t},n_{\text{f}}}}\,{\mathbf{H}}_{\text{sen}, n_{\text{f}}}^{\text{T}}\mathbf{x}_{n_{\text{f}}} + \mathbf{z}_{\text{sen}, n_{\text{f}}},
\end{equation}
\hspace{-0.5em}where ${\mathbf{H}}_{\text{sen}, n_{\text{f}}}\in\mathbb{C}^{N_{\text{t}}\times N_{\text{r}}}$ represents the multipath channel between the TX and the RX, and 
$\mathbf{z}_{\text{sen}, n_{\text{f}}}$ denotes the additive noise at the RX.
According to Fig. \ref{fig-model}, the sensing channel ${\mathbf{H}}_{\text{sen}, n_{\text{f}}}$ can be expressed as
\begin{equation}\label{eq-sensing-channel}
\begin{aligned}
{\mathbf{H}}_{\text{sen}, n_{\text{f}}} = & \ \mathbf{H}_{\text{si}, n_{\text{f}}} + \mathbf{H}_{\text{tx}\text{-}\text{roi}\text{-}\text{rx}, n_{\text{f}}} + \mathbf{H}_{\text{tx}\text{-}\text{ris}\text{-}\text{rx}, n_{\text{f}}} \\
& + \mathbf{H}_{\text{tx}\text{-}\text{roi}\text{-}\text{ris}\text{-}\text{rx}, n_{\text{f}}} + \mathbf{H}_{\text{tx}\text{-}\text{ris}\text{-}\text{roi}\text{-}\text{rx}, n_{\text{f}}} + {\mathbf{H}}_{\text{others}, n_{\text{f}}},
\end{aligned}
\end{equation}
where $\mathbf{H}_{\text{si}, n_{\text{f}}} = \beta \mathbf{H}_{\text{tx}\text{-}\text{rx}, n_{\text{f}}}$ denotes the quasi-static self-interference (SI) residual channel \cite{chen2018self}.
Here, $\mathbf{H}_{\text{tx}\text{-}\text{rx}, n_{\text{f}}}$ is the LOS path between the TX and the RX, and ${\beta\in[0,1]}$.
Typically, $\beta$ is a small value due to advanced SI cancellation techniques \cite{mehrotra2022degrees,ni2023receiver,zhang2015full}.
${\mathbf{H}}_{\text{others}, n_{\text{f}}}$ is the multipath components scattered by the UE or other random scatterers, considered disturbances to the sensing function.\footnote{
In applications like human posture recognition, the UE itself may be the target of sensing \cite{hu2020reconfigurable,yao2015rf,sasakawa2018human,zhao2023intelligent}. In such cases, channels related to the UE should be included in ${\mathbf{H}}_{\text{sen}, n_{\text{f}}}$.
However, this study considers a general scenario where the UE is not the target. Additional reflected or scattered paths may also be incorporated into ${\mathbf{H}}_{\text{sen}, n_{\text{f}}}$, depending on the system configuration.}
The other multipath components in \eqref{eq-sensing-channel} are defined similarly to those in \eqref{eq-commun-channel} and are summed incoherently due to the phase shifts caused by different transmission delays \cite{gonzalez2024integrated}.

The end-to-end channel ${\mathbf{H}}_{\text{sen}, n_{\text{f}}}$ can be estimated using the least squares (LS) algorithm \cite{tang2023joint}, leveraging $N_{\text{t}}$ received DL signals on the $n_{\text{f}}$-th subcarrier.
The transmit signals, denoted as $\mathbf{X}_{n_{\text{f}}} = [\mathbf{x}_{1, n_{\text{f}}}, \mathbf{x}_{2, n_{\text{f}}}, \ldots, \mathbf{x}_{N_{\text{t}}, n_{\text{f}}}]$, are known at the BS and can be designed using dedicated precoding techniques \cite{lu2024random}.\footnote{
The number of measurements used for estimating ${\mathbf{H}}_{\text{sen}, n_{\text{f}}}$ may be reduced to be much lower than $N_{\text{t}}$ by harnessing the sparse property of ${\mathbf{H}}_{\text{sen}, n_{\text{f}}}$ \cite{tang2023joint}. In this study, we take the simple LS algorithm as an example for analysis.}
The estimate of ${\mathbf{H}}_{\text{sen}, n_{\text{f}}}$ can be given as
\begin{equation}\label{eq-sensing-measurement}
\widehat{\mathbf{H}}_{\text{sen}, n_{\text{f}}} = {\mathbf{H}}_{\text{sen}, n_{\text{f}}} + \mathbf{Z}_{\text{sen}, n_{\text{f}}},
\end{equation}
where $\mathbf{Z}_{\text{sen}, n_{\text{f}}}$ is the estimation error resulting from the additive noise $\mathbf{z}_{\text{sen}, n_{\text{f}}}$ and the random disturbance $\mathbf{H}_{\text{others}, n_{\text{f}}}$.
By stacking the measurements, we obtain the 3D CSI data $\widehat{\boldsymbol{\mathcal{H}}}_{\text{sen}}\in\mathbb{C}^{N_{\text{t}}\times N_{\text{r}}\times N_{\text{f}}}$ used for target recognition, whose noise-free version is denoted as  $\boldsymbol{\mathcal{H}}_{\text{sen}}$.
No subpath extraction procedure is required in the proposed method, and the measurement $\widehat{\boldsymbol{\mathcal{H}}}_{\text{sen}}$ is directly used for target recognition.

\section{Protocol Design and Spectral Efficiency Analysis}

\subsection{Protocol Design}

We design the protocol for the integrated communication and recognition system based on the 5G new radio (NR) frame structure, considering its flexible uplink (UL)/DL switching capabilities \cite{ji2023networking} and the fast reconfiguration time of RIS phases \cite{li2020intelligent}. The RIS is assumed to assist in both communication and sensing. 
To improve communication performance, the RIS phases can be optimized to maximize the spectral efficiency (SE) or sum-rate.
According to \cite{huang2024single}, optimizing RIS phases based solely on the center subcarrier frequency can achieve nearly the same performance as optimization across multiple frequency points, especially in systems with small bandwidth and high-resolution RIS phase bits.
Therefore, we employ the algorithm proposed in \cite{wu2018intelligent} to determine the optimal RIS phase configuration, denoted as $\boldsymbol{\omega}_{\text{com}}$, which maximizes the SE at the center subcarrier.
The detailed optimization problem formulation is provided in Appendix \ref{appendix-optimization}.
Furthermore, the precoding of the transmitted signal $\mathbf{x}_{n_{\text{f}}}$ during the communication phase can be implemented using the maximum ratio transmission \cite{wu2018intelligent}.

To achieve target recognition during communication, we propose that the TX transmits DL signals, and the RIS phase shifts are configured according to the method described in Sec. \ref{sec-ris-phase-design} during the last $N_{\text{t}}$ symbol intervals in each frame, denoted as $\boldsymbol{\omega}_{\text{sen}, k}$, where $k=1, 2, \ldots, K$.
These $N_{\text{t}}$ symbols are received simultaneously by the UE and the RX, enabling communication and sensing, respectively.
The RX estimates $\widehat{\boldsymbol{\mathcal{H}}}_{\text{sen}}$ using the LS algorithm. 
Other communication processes are not affected by this method.
Varying the RIS phases with $K$ distinct configurations, the target label can be predicted.

\begin{figure}
    \centering
    \includegraphics[width=0.85\linewidth]{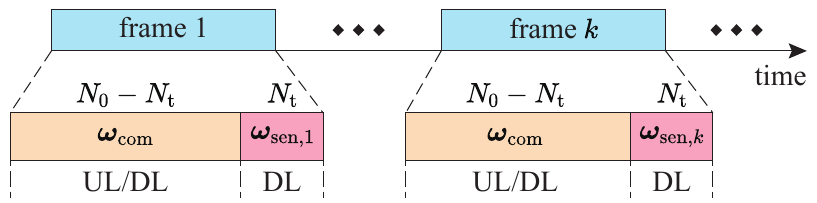}
    \captionsetup{font=footnotesize}
    \caption{The proposed protocol with time-division RIS configurations.} 
    \label{fig-protocol}
\end{figure}

The protocol is depicted in Fig. \ref{fig-protocol}, where $N_0 = 140\times 2^{\mu}$ represents the number of OFDM symbols in one frame, and  $\mu$ is the 5G NR numerology.
Since we use CSI measurements $\widehat{\boldsymbol{\mathcal{H}}}_{\text{sen}}$ directly for target recognition, the required number of measurements is expected to be significantly reduced compared to methods that first reconstruct the target image and then classify it \cite{saigre2022intelligent,del2020learned}, as demonstrated in Sec. \ref{sec-simu-first-imaging}. This allows for real-time recognition, even though intermittent sensing intervals are implemented to match the delay in RIS phase generation. Additionally, this ensures high communication rates within each frame.

\subsection{SE Analysis}

To evaluate the impact of sensing on communication performance, we calculate the SE of the proposed system.
We assume the system employs a comb-type pilot structure, estimating the DL communication channel $\mathbf{h}_{\text{com}, n_{\text{f}}}$ at each symbol interval \cite{wei2023carrier}.
Additionally, we assume that the positions of all elements in Fig. \ref{fig-model} remain static within a single frame.
Consequently, the channel $\mathbf{h}_{\text{com}, n_{\text{f}}}$ depends only on $\boldsymbol{\omega}$, written as $\mathbf{h}_{\text{com}, n_{\text{f}}}(\boldsymbol{\omega})$.
Symbolizing the noise variance as $\sigma^2_{\text{com}}$, the SE of DL communication with perfect CSI can be expressed as
\begin{equation}\label{eq-se1}
{\text{SE}}(\boldsymbol{\omega}) = \sum_{n_{\text{f}}=1}^{N_{\text{f}}}\log_{2}{\left(1+\frac{P_{\text{t}}\left\|\mathbf{h}_{\text{com}, n_{\text{f}}}(\boldsymbol{\omega})\right\|^{2}_2}{\sigma^2_{\text{com}}}\right)}.
\end{equation}
Based on the proposed protocol in Fig. \ref{fig-protocol}, the average SE is given by
\begin{equation}\label{eq-se2}
\overline{\text{SE}}_{\mu} = \frac{N_0-N_{\text{t}}}{N_0} {\text{SE}}(\boldsymbol{\omega}_{\text{com}}) + \frac{N_{\text{t}}}{N_0} {\text{SE}}(\boldsymbol{\omega}_{\text{sen}}),
\end{equation}
which varies with the 5G NR numerology $\mu$. Since the number of antennas $N_{\text{t}}$ is typically much smaller than the number of symbols $N_0$, the communication performance loss is considered negligible compared to ${\text{SE}}(\boldsymbol{\omega}_{\text{com}})$.

Moreover, the proposed protocol can be modified to speed up the sensing process by allocating more symbol intervals for sensing in one frame.
Consequently, multiple CSI measurements can be acquired with varying RIS phases, whereas the communication performance may be degraded.
The tradeoff between sensing and communication performances is analyzed in Sec. \ref{sec-simu-communication}.

\begin{figure*}
    \centering
    \includegraphics[width=0.9\linewidth]{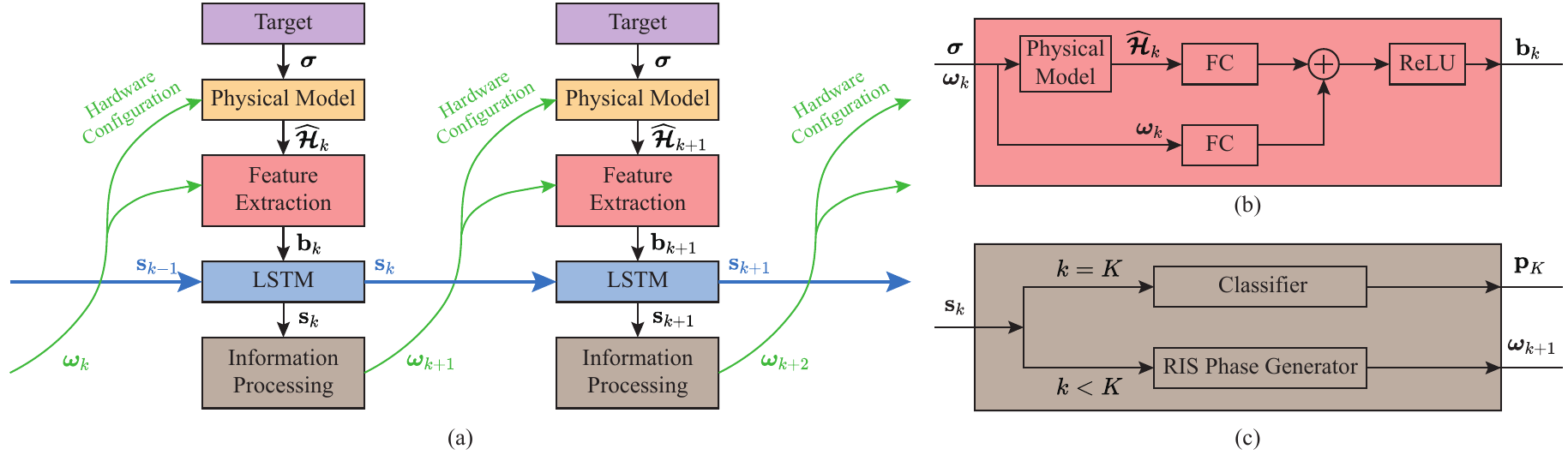}
    \captionsetup{font=footnotesize}
    \caption{(a) The proposed NN based on the LSTM architecture; (b) Structure of the physical model and feature extraction module; (c) Details of the information processing module.} 
    \label{fig-net}
\end{figure*}

\section{Recognizer with Adaptive RIS Phase Customization}
\label{sec-ris-phase-design}

In this section, we focus on designing RIS phase shifts to enhance sensing accuracy with limited measurements.
Inspired by the observation that the measurements acquired at previous moments contain valuable information about the target, we propose to customize RIS phases by iteratively utilizing the available information.
RIS configurations are adaptively tailored to the scene, the recognition task, the target being sensed, and RIS phase quantization constraint.
Furthermore, given the highly coupled properties between the RIS phase configuration $\boldsymbol{\omega}_{\text{sen}, k}$ and the NN parameters $\boldsymbol{\theta}$, their values are jointly learned through supervised learning by cooperating with the priorly known physical channel model.

\subsection{Overall Design of the NN}

Drawing on techniques from \cite{mnih2014recurrent,sohrabi2022active}, we design our NN based on a LSTM architecture, as illustrated in Fig. \ref{fig-net}(a).
At each moment $k$, corresponding to the $k$-th frame in Fig. \ref{fig-protocol}, the proposed NN merges the information from the $k$ obtained measurements and adjusts the RIS configuration for the subsequent $(k+1)$-th moment to gather the most relevant information for identifying the target class.
The newly generated RIS phase $\boldsymbol{\omega}_{\text{sen},k+1}$ is then applied to the RIS hardware, acquiring a new measurement for further analysis at the $(k+1)$-th moment.
Hence, the optimized RIS phases are specifically tailored for each target at each moment.
By collaborating with the physical channel model, our proposed method extensively utilizes the prior information about the scene.
Furthermore, the proposed NN aims to simultaneously optimize the system's hardware (i.e., RIS phases) and software (i.e., NN parameters) components, which are trained through supervised learning for the recognition task and considering RIS phase quantization constraints.
Employing the available information, the proposed NN can gradually recover the comprehensive information of the target's shape and scattering characteristics over time, thereby facilitating accurate recognition at the last moment.

\subsection{Key Modules of the NN}

In this subsection, we detail the key modules of the proposed NN architecture.

\textbf{Physical Model:}
This module is a reflection of the physical wave interactions, which projects the target image $\boldsymbol{\sigma}$ to the channel measurement $\widehat{\boldsymbol{\mathcal{H}}}_k$ (eliminating the subscript $(\cdot)_{\text{sen}}$) with the given RIS phase configuration $\boldsymbol{\omega}_k$.
Under the assumption of static cascaded channels, the noise-free multipath channel $\boldsymbol{\mathcal{H}}_{k}$ is the function of the target image $\boldsymbol{\sigma}$ and the RIS phase configuration $\boldsymbol{\omega}_k$.
The relationship is given as
\begin{equation}\label{eq-physical-model}
\boldsymbol{\mathcal{H}}_{k} = f_{\text{phy}} (\boldsymbol{\sigma}, \boldsymbol{\omega}_{k}),
\end{equation}
where $f_{\text{phy}}$ represents the physical channel model, whose detailed form is provided in Appendix \ref{appendix-channel}.
$f_{\text{phy}}$ includes no learnable parameters, since the relationship shown in \eqref{eq-physical-model} is assumed to be priorly known with the available locations of the TX, RX, RIS, and ROI.

The proposed NN is expected to optimize the RIS phases while accounting for RIS hardware impairments, such as frequency selectivity, directional selectivity, and amplitude-phase coupling.\footnote{We do not consider these impairments in this study, and it is in our future work.} In practice, $f_{\text{phy}}$ can also be approximated by a well-trained NN \cite{li2020intelligent,yuan2024neural}, which integrates physical and non-ideal factors in the system model.

\textbf{Feature Extraction:}
This module extracts the information involved in the raw measurement $\widehat{\boldsymbol{\mathcal{H}}}_k$ to a feature vector $\mathbf{b}_k\in \mathbb{R}^{B_1}$, which is subsequently input to the LSTM module for information fusion and processing.
Here, $B_1$ is the output dimension of the FC layers.
Inspired by \cite{mnih2014recurrent}, we simultaneously input the RIS phase shift $\boldsymbol{\omega}_{k}$ and the measurement $\widehat{\boldsymbol{\mathcal{H}}}_k$ to this module, guiding the NN to extract information about the target $\boldsymbol{\sigma}$ in the ROI.
This module can be formulated as
\begin{equation}
{\mathbf{b}}_{k} = f^{\boldsymbol{\theta}_{1}}_{\text{fea}} (\widehat{\boldsymbol{\mathcal{H}}}_k, \boldsymbol{\omega}_{k}),
\end{equation}
where $\boldsymbol{\theta}_{1}$ is the learnable parameters.
Specifically, the real and imaginary parts of the complex CSI measurement $\widehat{\boldsymbol{\mathcal{H}}}_k\in\mathbb{C}^{N_{\text{t}}\times N_{\text{r}}\times N_{\text{f}}}$ are flattened into a real vector ${\mathbf{h}}_k\in\mathbb{C}^{2N_{\text{t}}N_{\text{r}}N_{\text{f}}}$.
Then, ${\mathbf{h}}_k$ is input to the FC layer with the input node number $2N_{\text{t}}N_{\text{r}}N_{\text{f}}$.
The phase information of $\boldsymbol{\omega}_{k}$ is input to another independent FC layer, and the outputs of the FC layers are summed and activated by the rectified linear unit (ReLU).
As a result, the target information across multiple frequency points is extracted to assist in customizing the RIS phases.
The physical model and feature extraction module form a residual structure, as shown in Fig. \ref{fig-net}(b).

\textbf{LSTM}\footnote{The LSTM layer can be replaced by other RNN structures, such as the gated recurrent unit (GRU).}\textbf{:}
This is the core module of the proposed NN.
In this study, we propose to employ historical measurements to instruct RIS phase design.
However, the feature vector $\mathbf{b}_k$ output by the feature extraction module accumulates as the RIS phases change successively.
Hence, we employ the LSTM module to iteratively extract and fuse the information lying in the feature vector $\mathbf{b}_k$ and the state vector $\mathbf{s}_{k-1}$, given as
\begin{equation}
{\mathbf{s}}_{k} = f^{\boldsymbol{\theta}_{2}}_{\text{lstm}} (\mathbf{b}_k, \mathbf{s}_{k-1}),
\end{equation}
where $\boldsymbol{\theta}_{2}$ is the learnable parameters.
${\mathbf{s}}_{k}\in\mathbb{R}^{B_2}$ denotes the state vector output by the LSTM module at the $k$-th moment, summarizing the target information lying in previous measurements to a constant-dimension vector.
The input and output dimensions of the LSTM module are $B_1$ and $B_2$, respectively.

With the accumulation of the measurements, rich information about the target is embedded into $\mathbf{s}_{k}$, which can be projected to one of the available categories.
Moreover, $\mathbf{s}_{k}$ may also reflect the absence of certain information, which is essential for the system to make solid predictions, guiding the RIS phase design.
Consequently, the state vector $\mathbf{s}_{k}$ is transferred to the information processing module for class prediction or RIS phase customization, as well as the LSTM module at the next moment for information fusion.

\textbf{Information Processing:}
This module intakes the state vector $\mathbf{s}_{k}$ and makes actions with the extracted information contained in the acquired $k$ measurements.
Specifically, $\mathbf{s}_{k}$ is input to two sub-modules, the classifier and the RIS phase generator, as depicted in Fig. \ref{fig-net}(c).
The classifier only works when $k=K$ and outputs the probabilities that the target belongs to each category, denoted as $\mathbf{p}_K \in \mathbb{R}^{N_{\text{c}}\times 1}$, where $N_{\text{c}}$ is the number of possible categories.
The RIS phase generator works at each moment when $k<K$, adaptively generating the best RIS configuration $\boldsymbol{\omega}_{k+1}$ for the next moment, which is configured to the RIS hardware at the last $N_{\text{t}}$ symbol intervals of the $(k+1)$-th frame.

The two sub-modules are composed of two independent FC layers, which can be described as
\begin{equation}
{\mathbf{p}}_{K} = f^{\boldsymbol{\theta}_{3}}_{\text{cla}} (\mathbf{s}_{K}),\quad {\boldsymbol{\omega}}_{k+1} = f^{\boldsymbol{\theta}_{4}}_{\text{pha}} (\mathbf{s}_{k}),
\end{equation}
where $\boldsymbol{\theta}_{3}$ and $\boldsymbol{\theta}_{4}$ are the learnable parameters of the classifier and the RIS phase generator, respectively.
We employ the SoftMax function for the activation of the classifier, ensuring that the sum of the elements in ${\mathbf{p}}_{K}$ equals one.
Alternatively, the output of the RIS phase generator is activated by the hyperbolic tangent (Tanh) function and multiplied by $\pi$ to force the optimized RIS phases to be restricted in the range of $[-\pi, \pi]$.
Since the state vector $\mathbf{s}_{k}$ is unique for each target at each moment, $f^{\boldsymbol{\theta}_{4}}_{\text{pha}}$ generates different RIS phase shifts, which are called by {\bf RIS phase customization} in the proposed NN.
However, the first RIS configuration $\boldsymbol{\omega}_1$ is the same for each target, which can also be learned by the NN by integrating it as part of the trainable parameters, as studied in \cite{del2020learned}.
Therefore, $\boldsymbol{\omega}_1$ is explicitly defined and optimized, whereas the design of $\{\boldsymbol{\omega}_k\}_{k=2}^K$ is implicitly represented by the NN parameters $\boldsymbol{\theta}_{4}$ and should be derived through NN inference, which is similar to the concept of implicit neural representation \cite{mildenhall2021nerf}.
Note that the proposed method can be easily extended to multi-RIS scenarios by modifying the physical channel model and the RIS phase generator.

\subsection{Training and Implementation}
\label{sec-net-training}

We initially consider NN training with continuous RIS phase shifts in this subsection, where the NN $f_{\boldsymbol{\theta}}$ can be trained using the gradient descent algorithm.
The cross-entropy classification loss function is used in our study, given as
\begin{equation}\label{eq-cross-entropy}
L_{\text{CE}} = -\frac{1}{M}\sum_{m=1}^M\log({p}_{K,c_m}),
\end{equation}
where ${c_m\in \{1, 2, \ldots, N_{\text{c}}\}}$ is the index of the true target label for the $m$-th training data.
${p}_{K,c_m}$ denotes the $c_m$-th element in vector $\mathbf{p}_{K}$.
$M$ is the number of training samples.

During training, $L_{\text{CE}}$ is minimized to optimize the NN parameters $\boldsymbol{\theta}=\{\boldsymbol{\theta}_1, \boldsymbol{\theta}_2, \boldsymbol{\theta}_3, \boldsymbol{\theta}_4, \boldsymbol{\omega}_1\}$. Gradients are computed and backpropagated through the different modules of the network to update these parameters.
Additionally, during the training phase, the target image $\boldsymbol{\sigma}$ is required to generate CSI measurements using the physical model. However, during actual NN implementation, the target image is no longer needed. Instead, only the CSI measurements, which are obtained from channel estimation at the RX, are used as input to the NN.

Note that the classifier is directly responsible for target recognition in the proposed NN, using the final output of the LSTM unit. As a result, a well-trained model can adapt quickly to different noise levels in the environment, requiring only the classifier parameters to be retrained with new data, thus making the system highly adaptable.

\begin{remark}
The proposed NN is closely related to the RL framework \cite{mnih2014recurrent,hu2021metasensing}, where the BS or RIS can be considered an intelligent agent, which interacts with the radio propagation environment through the action defined by the generated RIS phases.
Furthermore, the minus cross-entropy loss can be considered as the reward function.
However, we do not employ RL techniques to train the proposed NN, since it is an end-to-end RNN architecture, and all its components are differentiable.
Thus, the impact of each generated RIS phase configuration can be directly revealed and tuned by the backpropagated gradients during supervised learning.
Additionally, the next subsection further extends the application scope of the gradient descent algorithm to discrete RIS phase scenarios.
Furthermore, supervised learning is anticipated to converge faster and possess higher data efficiency when compared to RL techniques \cite{sohrabi2022active}.
Simulation results in Sec. \ref{sec-simulation} demonstrate that the trained NN with supervised learning achieves high sensing performance.
\end{remark}

\subsection{Implementation of Discrete RIS Phase Shifts}
\label{sec-discrete-ris-phase}

Discrete RIS phases are practical in hardware implementation \cite{huang2024ris}.
The basic approach for managing discrete RIS phases involves quantifying the optimized continuous configurations to discrete values.
However, considering the strong coupling between RIS phases and NN parameters \cite{hu2021metasensing}, the discretization may result in mismatches between them.
Although the classifier can be retrained with $\mathbf{s}_K$ obtained from measurements of discretized RIS phases, they may still produce degraded sensing performance.
Thus, the NN should be specifically designed to optimize discrete RIS phases.
However, the traditional gradient descent algorithm is originally not suitable for the optimization of discrete parameters, due to their discontinuous gradients.
Although RL techniques have been exploited for discrete RIS phase design \cite{hu2021metasensing}, they typically deserve high training difficulty, since the available searching space is significantly large for the proposed method with numerous RIS elements and multiple distinct configurations.

\begin{figure}
    \centering
    \includegraphics[width=\linewidth]{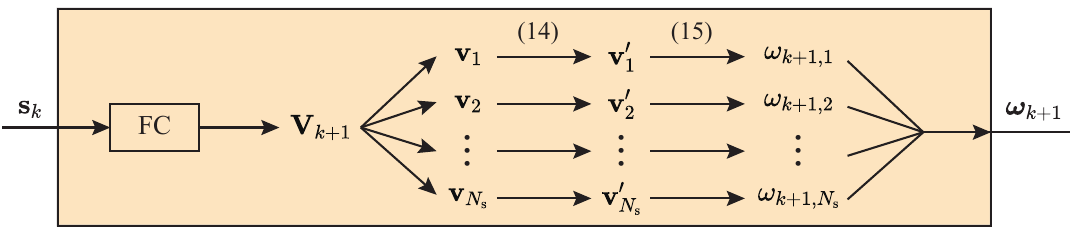}
    \captionsetup{font=footnotesize}
    \caption{The proposed RIS phase generator under quantization constraints.}
    \label{fig-discrete-ris-phase}
\end{figure}

In this study, we introduce a temperature parameter \cite{chakrabarti2016learning} to train the proposed NN with discrete RIS phase shifts, addressing a limitation not considered in \cite{sohrabi2022active}.
The training strategy starts with continuous RIS phase values and progressively forces their distributions to become increasingly discrete over the training course, ending with effectively discretized RIS phase values.
We assume that the RIS phases are $b$-bit quantified and can only choose $Q=2^b$ discrete values from the set $\mathcal{O} = \{\frac{2\pi}{Q}, \frac{4\pi}{Q}, \ldots, \frac{2(Q-1)\pi}{Q}, 2\pi\}$.
Training the NN with discrete RIS phase values requires a redesign of the RIS phase generator, since conventional FC layers apply no constraints on the output RIS phases.
Alternatively, we propose to guide the FC layer in the RIS phase generator to produce a weighting matrix $\mathbf{V}_{k+1}\in\mathbb{R}^{N_{\text{s}}\times Q}$, whose $(n_{\text{s}}, q)$-th element denotes the weight for the $n_{\text{s}}$-th RIS element to choose the $q$-th possible discrete value in $\mathcal{O}$ at the $(k+1)$-th moment, as depicted in Fig. \ref{fig-discrete-ris-phase}.
We define ${\omega}_{k+1,n_{\text{s}}}$ as the RIS phase value of the $n_{\text{s}}$-th RIS element optimized for the $(k+1)$-th moment, whose weighting vector is given as $\mathbf{v}_{n_{\text{s}}} = [v_{n_{\text{s}},1}, v_{n_{\text{s}},2}, \ldots, v_{n_{\text{s}},Q}]^{\text{T}}\in\mathbb{C}^{Q\times 1}$.
Moreover, we define a temperature parameter $\alpha(t)$, which increases with the training procedure, given as
\begin{equation}
\alpha(t) = 1 + (\phi \cdot t)^2,
\end{equation}
where $\phi$ is a tunable hyper-parameter that controls the increase of $\alpha(t)$.
$t$ is the index of the training batch and is recorded at the last batch for model tests.
At the $t$-th batch, we define
\begin{equation}\label{eq-vns}
\mathbf{v}_{n_{\text{s}}}' = f_{\text{softmax}}\left(\alpha(t)\left|\mathbf{v}_{n_{\text{s}}}\right|\right),
\end{equation}
where $|\mathbf{v}_{n_{\text{s}}}| = \left[\left|v_{n_{\text{s}},1}\right|, \left|v_{n_{\text{s}},2}\right|, \ldots, \left|v_{n_{\text{s}},Q}\right|\right]^{\text{T}}$, and $f_{\text{softmax}}$ denotes the SoftMax function.
The $q$-th element of $\mathbf{v}_{n_{\text{s}}}'$, ${v}_{n_{\text{s}},q}'$, can be considered as the probability for ${\omega}_{k+1,n_{\text{s}}}$ to choose the $q$-th value in $\mathcal{O}$, i.e., $2\pi q/Q$.
Consequently, the optimized RIS phase value for ${\omega}_{k+1,n_{\text{s}}}$ can be calculated by
\begin{equation}
{\omega}_{k+1,n_{\text{s}}} = \sum_{q=1}^Q {v}_{n_{\text{s}},q}'\frac{2\pi q}{Q}.
\end{equation}

During the NN training phase, the weighting matrix $\mathbf{V}_{k+1}$ can be considered a hidden layer of the proposed NN and optimized by the backpropagation algorithm.
The SoftMax function $f_{\text{softmax}}$ and the gradually increasing temperature parameter $\alpha(t)$ forces the vector $\mathbf{v}_{n_{\text{s}}}'$ to progressively contain only a single element as one whereas other elements are zero.
Consequently, the RIS phase ${\omega}_{k+1,n_{\text{s}}}$ ultimately converges to one of the possible values in $\mathcal{O}$.
Therefore, we have redesigned the RIS phase generator under quantization constraints, allowing for gradient calculation and backpropagation.

\begin{remark}
The proposed NN presents significant advancements over prior studies by integrating RIS configurations and NN parameters for joint optimization, diverging from the separate approaches used in previous works \cite{hu2020reconfigurable,zhao2023intelligent,li2019machine}.
Our NN customizes RIS phase shifts for individual targets, utilizing prior knowledge of the scene, task, target, and quantization information to enhance sensing performance---an approach not explored in earlier studies \cite{hu2021metasensing,del2020learned}.
Moreover, the RIS phase designs are implicitly represented within the NN parameters, allowing the network to adaptively generate RIS phases during the sensing process for each target. This differs significantly from the LISP method \cite{del2020learned}, which optimizes RIS configurations for all targets as explicitly trainable parameters of the NN.
Additionally, the dynamic mode of RIS phase generation, combined with the LSTM architecture, allows for the design of flexible sensing durations, which will be discussed in the next section and further contribute to improved sensing performance. Despite the complexity of the proposed NN and the time required to generate RIS configurations, our method ensures efficiency by employing the protocol shown in Fig. \ref{fig-protocol}, which incorporates designed sensing intervals.
\end{remark}

\section{Enhanced Recognizer with Dynamic Sensing Durations}
\label{sec-decision}

Prior studies have designed NNs with a fixed number of CSI measurements \cite{hu2020reconfigurable,zhao2023intelligent,li2019machine,hu2021metasensing,del2020learned,sohrabi2022active}.
However, we note that varying targets may possess unique difficulties for classification: ``easy'' targets that possess clear category features can be recognized with short sensing durations, whereas ``hard'' samples with vague category features request large numbers of CSI measurements.
Hence, employing a fixed number of measurements may be redundant for these ``easy'' targets and inadequate for ``hard'' instances.
In this section, we aim to customize the sensing durations for different targets, terminating the sensing procedure as long as adequate information has been acquired from the previous measurements.
This is crucial for accelerating the sensing procedure.
Specifically, we first enhance the NN proposed in Sec. \ref{sec-ris-phase-design}.
Then, we detail the training strategy to achieve superior sensing performance.

\subsection{NN Enhancement with a Learned Decisioner}

Referring to advanced studies about dynamic NN structures \cite{liu2023resource,xu2023lgvit,wang2020glance,li2017dynamic}, a decisioner that controls the continuation and termination of the sensing process can be introduced to the NN proposed in Sec. \ref{sec-ris-phase-design}.
Considering that we have utilized an LSTM architecture where each iteration of the LSTM unit corresponds to the acquisition and processing of one CSI measurement, the sensing procedure can be flexibly terminated at any moment.
Inspired by conventional techniques in the CV field \cite{mnih2014recurrent,li2017dynamic}, we propose to learn a policy that adaptively stops the sensing process according to the acquired target information, which is called by {\bf sensing duration customization}.

\begin{figure}
    \centering
    \includegraphics[width=0.6\linewidth]{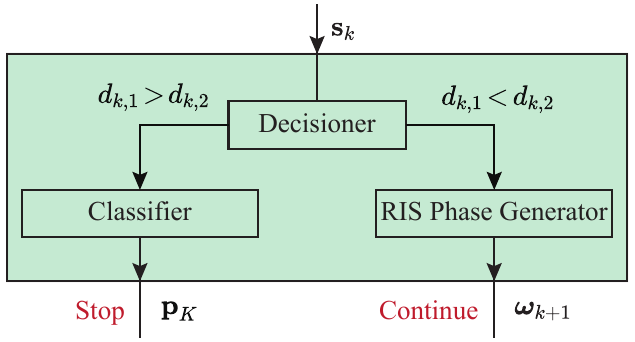}
    \captionsetup{font=footnotesize}
    \caption{The modified information processing module with a decisioner.}
    \label{fig-decisioner}
\end{figure}

Specifically, we modify the NN structure in Sec. \ref{sec-ris-phase-design} by adding a decisioner into the information processing module.
The decisioner consists of an FC layer activated by the SoftMax function, which intakes the state vector $\mathbf{s}_k$ of the LSTM module and outputs a two-dimensional vector $\mathbf{d}_k \in\mathbb{C}^{2\times1}$, whose elements, $d_{k,1}$ and $d_{k,2}$, represent the probabilities to stop or continue the measurement acquisition process at the $k$-th moment, respectively.
The decisioner can be expressed as
\begin{equation}
\mathbf{d}_k = f^{\boldsymbol{\theta}_{5}}_{\text{dec}} (\mathbf{s}_{k}),
\end{equation}
where $\boldsymbol{\theta}_{5}$ is the learnable parameters.
The decisioner works at every moment to terminate the sensing process timely as long as $\mathbf{s}_k$ has sufficiently pictured the category features of the target.
Different from the classifier in Sec. \ref{sec-ris-phase-design} that only works when $k=K$, the classifier in this section should output the probability vector $\mathbf{p}_k$ at any moment when the decisioner concludes to stop the sensing procedure, i.e., when $d_{k,1}>d_{k,2}$.
Alternatively, the RIS phase generator should tailor the RIS phases for the next moment if $d_{k,1}<d_{k,2}$.
The operation mechanism of the modified information processing module is depicted in Fig. \ref{fig-decisioner}, which adaptively customizes both the RIS phase configurations and sensing duration for the target being sensed.
Other parts of the NN are the same as that in Sec. \ref{sec-ris-phase-design}.

\subsection{Training}

In the proposed NN architecture, automatically terminating the sensing procedure leads to different iteration numbers of the LSTM module, i.e., distinct NN structures.
Hence, the model structure ${S}$ is dynamic, depending on both the target being sensed and the NN parameters $\boldsymbol{\theta}'=\{\boldsymbol{\theta}_1, \boldsymbol{\theta}_2, \boldsymbol{\theta}_3, \boldsymbol{\theta}_4, \boldsymbol{\theta}_5, \boldsymbol{\omega}_1\}$.
Given a target in the ROI with its scattering characteristics represented by the radio image $\boldsymbol{\sigma}$, we denote the probability of choosing an NN structure ${S}_k$ as $P({S}_k|\boldsymbol{\sigma}, \boldsymbol{\theta}')$.
Here, $S_k$ denotes the NN structure that the model is terminated at the $k$-th moment, and only $k$ CSI measurements are utilized for final class prediction.
Furthermore, the cross-entropy classification loss corresponding to the structure ${S_k}$ is symbolized by $L_{S_k}(\boldsymbol{\sigma}, \boldsymbol{\theta}')$.
Thus, the overall expected loss for the considered target can be formulated as
\begin{equation}\label{eq-loss-expected}
L_{\text{exp}} = \mathbb{E}_\mathcal{S}\left[L_\mathcal{S}(\boldsymbol{\sigma}, \boldsymbol{\theta}')\right] = \sum_{k=1}^{K_{\text{m}}} P({{S_k}}|\boldsymbol{\sigma}, \boldsymbol{\theta}')L_{S_k}(\boldsymbol{\sigma}, \boldsymbol{\theta}'),
\end{equation}
where $\mathcal{S}=\{S_1, S_2, \ldots, S_{K_{\text{m}}}\}$ denotes the set of available model structures, and $\mathbb{E}_\mathcal{S}[\cdot]$ calculates the expectation over $\mathcal{S}$.
$K_{\text{m}}$ is the maximum number of LSTM iterations, which is defined manually according to the available time resources.

According to \cite{li2017dynamic}, training an NN with dynamic durations requires the NN to explore structures and parameters that achieve the minimum expected loss $L_{\text{exp}}$.
RL techniques such as the REINFORCE algorithm \cite{sutton1999policy} can be employed for the optimization.
However, we note that the number of available model structures is finite in the proposed NN with $\text{card}(\mathcal{S})=K_{\text{m}}$, which may significantly degrade the NN training difficulty.
Here, $\text{card}(\mathcal{S})$ denotes the number of elements in the set $\mathcal{S}$.
Therefore, the model structure sampling operation in traditional RL techniques can be released, and the expected loss $L_{\text{exp}}$ can be directly calculated by traversing all the $K_{\text{m}}$ possible structures with their possibilities and cross-entropy losses.

The probability for choosing the $k^*$-th NN structure can be defined by the output of the decisioner, given as
\begin{equation}
P({S}_{k^*}|\boldsymbol{\sigma}, \boldsymbol{\theta}') = \left\{\begin{array}{ll}
d_{1,1}, & k^* = 1, \\[6pt]
d_{k^*,1}\prod_{k=1}^{k^*-1}d_{k,2}, & 1 < k^* < K_{\text{m}}, \\[6pt]
\prod_{k=1}^{K_{\text{m}}-1}d_{k,2}, & k^* = K_{\text{m}},
\end{array}\right.
\end{equation}
where $k^* = 1, 2, \ldots, K_{\text{m}}$, and $d_{k^*,1} + d_{k^*,2} = 1$.
Moreover, the cross-entropy loss can be derived based on \eqref{eq-cross-entropy}.
Consequently, the gradient of $L_{\text{exp}}$ can be computed based on \eqref{eq-loss-expected} and the backpropagation algorithm.

Note that the NN is not terminated at intermediate steps during the training phase, because the possibilities and losses stopped at each moment should be calculated for the expected loss $L_{\text{exp}}$.
Alternatively, only one NN structure is chosen during NN implementation, according to the termination mechanism depicted in Fig. \ref{fig-decisioner}.
Furthermore, we require that the NN can be trained with a tendency to employ large or small amounts of measurements.
Thus, a penalty factor $\gamma$ is introduced to the loss function, and \eqref{eq-loss-expected} can be rearranged as
\begin{equation}\label{eq-loss-expected2}
L_{\text{exp}}(\gamma) = \sum_{k=1}^{K_{\text{m}}} \gamma^k P(S_k|\boldsymbol{\sigma}, \boldsymbol{\theta}')L_{S_k}(\boldsymbol{\sigma}, \boldsymbol{\theta}'),
\end{equation}
where $\gamma > 1$ penalizes the NN structures that employ large amounts of measurements.
Degrading or increasing $\gamma$ are anticipated to terminate the sensing procedure with large or small amounts of measurements, respectively.

However, directly training the NN from scratch is an extremely hard task, since three functions, i.e., classification, RIS phase generation, and decision-making, are simultaneously learned.
Thus, we address this problem by employing the curriculum learning strategy \cite{bengio2009curriculum}.
Specifically, we first train an NN proposed in Sec. \ref{sec-ris-phase-design} with no decisioner.
Different from the training strategy described in Sec. \ref{sec-net-training}, we propose to force intermediate supervision during training, since the proposed model in this section may terminate at any moment.
Hence, the classifier outputs the predicted probabilities at each moment, and the loss function is the average cross-entropy classification loss over $K_{\text{m}}$ moments, given as
\begin{equation}
L_{\text{CE}}' = \frac{1}{K_{\text{m}}}\sum_{k=1}^{K_{\text{m}}} L_{S_k}(\boldsymbol{\sigma}, \boldsymbol{\theta}),
\end{equation}
where $\boldsymbol{\theta}$ is the trainable parameters with no decisioner.
Then, the pre-trained model is utilized to initialize the NN with a decisioner, and \eqref{eq-loss-expected2} is exploited to optimize $\boldsymbol{\theta}_5$ and fine-tune $\boldsymbol{\theta}$.
This strategy helps the model to converge to a better local optimum than training from scratch \cite{li2017dynamic}.

\section{Numerical Results}
\label{sec-simulation}

\subsection{Experimental Settings}

\subsubsection{Simulation Scenario}
We consider the simulation scenario depicted in Fig. \ref{fig-model}.
The center subcarrier frequency is set to $f_0=3$ GHz \cite{hu2021metasensing,li2020intelligent}.
The BS is located at $[30\lambda_0, 50\lambda_0, 50\lambda_0]^{\text{T}}$.
The RIS is deployed in the yOz plane with its center location at $[0, 0, 0]^{\text{T}}$.
RIS element size is $\xi_{\text{s}} = \lambda_0/2$.
We consider a two-dimensional ROI centered at $[D, 0, 0]^{\text{T}}$, which lies in the yOz plane and simplifies the occlusion effects between voxels in three-dimensional space \cite{huang2024ris}.
The UE location is $[30\lambda_0, -50\lambda_0, 0]^{\text{T}}$, and the received noise power at the UE and the RX is set to -80 dBm \cite{yang2020fast}.
The antenna gains $G_{\text{t}}$, $G_{\text{r}}$, and $G_{\text{ue}}$ are set to 2 according to \cite{sharma2022mimo}.

\subsubsection{Dataset Generation}
We employ the MNIST dataset with $N_{\text{c}}=10$ to simulate the target in the ROI \cite{del2020learned}, where $M = 60,000$ training data and 10,000 testing data are used.
The pictures in the dataset are padded and transformed into gray images with the size of $N_{\text{i}}=30\times 30$, whose pixel values are subsequently normalized to the range $[0, 4\pi A^2/\lambda_0^2]$, representing the RCS of the voxel with the area of $A=\lambda_0\times\lambda_0$ \cite{huang2024ris}.
Hence, we obtain the ROI images with the resolution of $\lambda_0$, and the numbers in these pictures are assumed to be the targets being sensed.
The normalized pixel values represent the scattering characteristics of the corresponding voxels in the ROI, and the CSI measurements can be generated with the physical channel model.

\begin{figure}[t]
    \centering
    \includegraphics[width=0.85\linewidth]{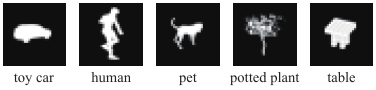}
    \captionsetup{font=footnotesize}
    \caption{Example images of the synthesized dataset.}
    \label{fig-dataset}
\end{figure}

To further demonstrate the effectiveness of the proposed method, we also utilize the Fashion-MNIST dataset \cite{xiao2017fashion} and a synthesized dataset. The synthesized dataset contains 6,000 images of five categories of everyday objects, originating from publicly available segmentation datasets on Kaggle (toy cars, humans, pets, and potted plants) or randomly generated using MATLAB (tables).
Example images of these five categories are shown in Fig. \ref{fig-dataset}, and the complete synthesized dataset can be found at \url{https://github.com/kiwi1944/CRISense}. Of the synthesized images, 5,000 are used for NN training, and the remaining images are used for testing. The preprocessing steps for the Fashion-MNIST and synthesized datasets follow the same process as that of the MNIST dataset.
In practical applications, other domain-specific datasets can be used to train the model \cite{del2020learned,li2020intelligent}.
Radio images used for NN training can be obtained through wireless imaging techniques \cite{zhu2016frequency,huang2024fourier,huang2024ris} or approximated by segmented binary optical images \cite{li2020intelligent,wang2024dreamer}.

\subsubsection{Training Details}
The NN training configurations include a batch size of 128, a total of 200 training epochs, and an initial learning rate of $10^{-3}$.
The validation set occupies 10\% of the training data.
The learning rate decays by 50\% if the validation accuracy is not improved for 5 epochs, and 20 epochs with no improvements terminate the training procedure.
The LSTM module contains a layer with $B_2 = 256$ hidden units, and its input size $B_1 = 256$.
Unless otherwise specified, all the FC layers consist of a single hidden layer with 256 neurons, with batch normalization and ReLU activation function applied after each layer.
The NN parameters are optimized with the Adam algorithm on an Nvidia 3090 GPU using the PyTorch platform.

In the next subsection, we evaluate the proposed methods through simulations, where the correct prediction rate $\eta$ on the testing dataset is taken as the performance evaluation metric.
From Sec. \ref{sec-simu-new} to Sec. \ref{sec-simu-interpretation}, we evaluate the performance of the proposed method in Sec. \ref{sec-ris-phase-design}, where Sec. \ref{sec-simu-new} discusses the basic settings of the system, and discrete RIS phase shifts are utilized in Sec. \ref{sec-simu-discrete}.
Then, the decisioner proposed in Sec. \ref{sec-decision} is validated in Secs. \ref{sec-simu-decisioner} and \ref{sec-simu-penalty}.

\subsection{Results and Discussions}

\subsubsection{Discussions of System Settings}
\label{sec-simu-new}

We evaluate the effects of various system parameters, such as the number of subcarriers, varying antenna numbers, different datasets, and various SI levels:

\begin{itemize}

\begin{figure}[t]
    \centering
    \includegraphics[width=0.78\linewidth]{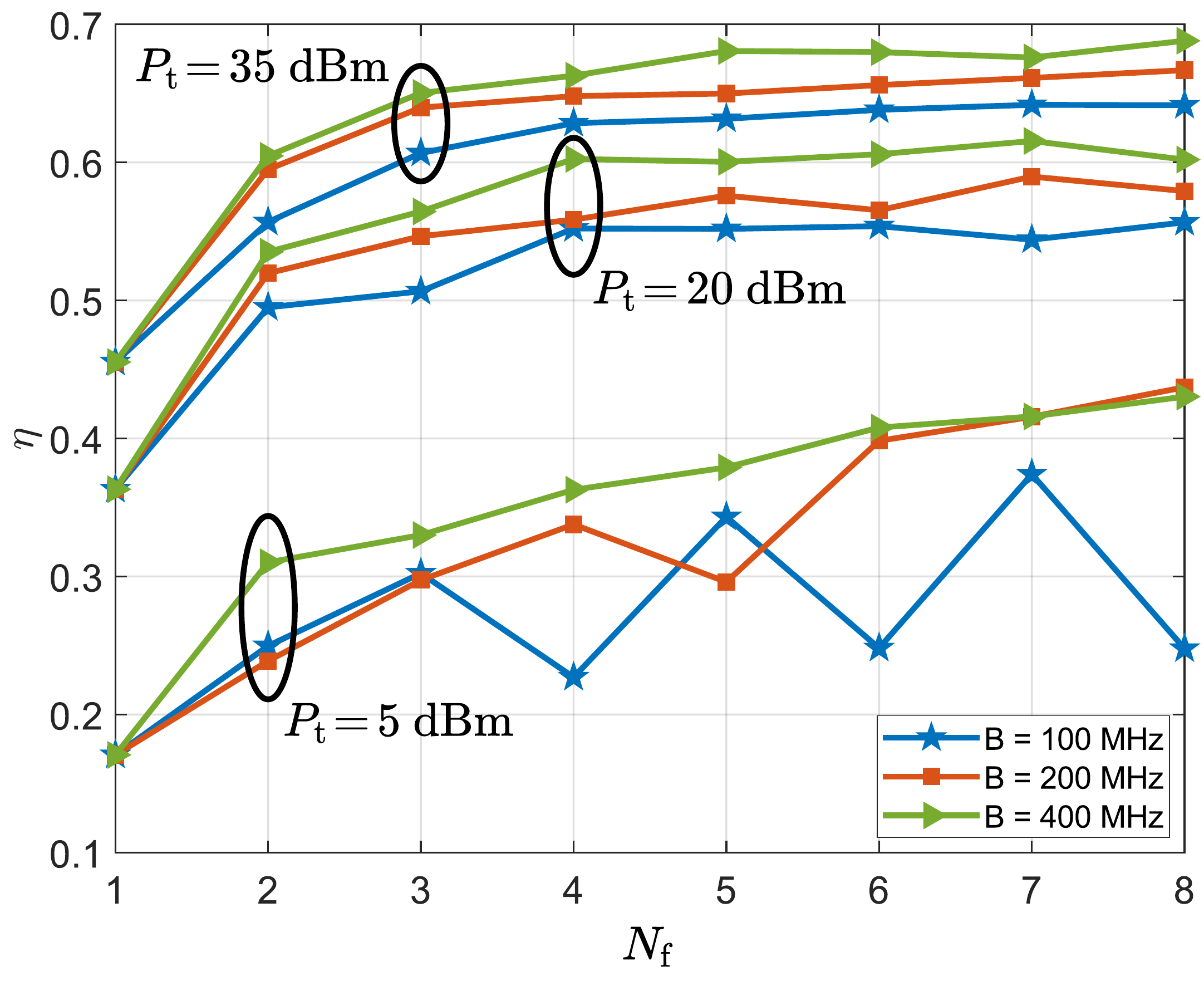}
    \captionsetup{font=footnotesize}
    \caption{Recognition accuracy $\eta$ with varying $N_{\text{f}}$, $B$ and $P_{\text{t}}$.}
    \label{fig-result-subcarrier}
\end{figure}

\item \textbf{Subcarrier Numbers and Bandwidths:}
We first investigate the effects of the number of utilized subcarriers $N_{\text{f}}$ and the bandwidth $B$ on sensing performance while maintaining a constant total transmit power $P_{\text{t}}$.
As a result, the power allocated to each subcarrier decreases with an increasing number of subcarriers, calculated as $P_{0} = P_{\text{t}} / N_{\text{f}}$.
The simulation setup includes a RIS of size $20\times 20$, a measurement count of $K=3$, and single antennas at both the TX and RX, i.e.,  $N_{\text{t}}=N_{\text{r}}=1$. For NN training, half of the MNIST dataset is utilized.
The simulation results, presented in Fig. \ref{fig-result-subcarrier}, demonstrate that the target recognition accuracy $\eta$ improves as more subcarriers are used for sensing, despite the reduction in signal-to-noise-ratio (SNR) on individual subcarriers. This performance improvement can be attributed to the increased number of CSI measurements, which provide richer information about the target and may partially compensate for the SNR degradation.
Additionally, when the subcarrier number $N_{\text{f}}$ is fixed, increasing the bandwidth $B$ further enhances sensing performance. This improvement is due to the expanded subcarrier spacing $\Delta f$, which captures more significant scattering characteristics of the target. However, at low transmit power levels (e.g., 5 dBm), sensing performance deteriorates. The relationship between $\eta$ and $N_{\text{f}}$ also shows more fluctuations, particularly when $B$ is small. This instability suggests that high noise levels introduce greater challenges for NN training, potentially limiting sensing accuracy.

\begin{table}[t]
    \renewcommand{\arraystretch}{1.4}
    \centering
    \fontsize{8}{8}\selectfont
    \captionsetup{font=small}
    \captionof{table}{Recognition accuracy with varying antenna numbers on different datasets.}\label{tab-antenna}
    \begin{threeparttable}
        \begin{tabular}{cccccc}
            \specialrule{1pt}{0pt}{-1pt}
            \xrowht{10pt} $N_{\text{tr}}$ & 1 & 2 & 3 & 4 & 5 \\
            \hline
            MNIST & 0.4876 & 0.6510 & 0.6553 & 0.6598 & \textbf{0.6608} \\
            Fashion-MNIST & 0.6414 & 0.7447 & 0.7456 & 0.7525 & \textbf{0.7594} \\
            Synthesized dataset & 0.8810 & 0.9350 & 0.9400 & 0.9570 & \textbf{0.9650} \\
            \specialrule{1pt}{0pt}{0pt}
        \end{tabular}
    \end{threeparttable}
\end{table}

\item \textbf{Antenna Numbers and Datasets:}
Next, we investigate the impact of varying antenna numbers with a $20\times 20$ RIS and $K=3$ RIS configurations.
For simplicity, we assume $N_{\text{t}}=N_{\text{r}}=N_{\text{tr}}$, and $N_{\text{f}}=1$. Table \ref{tab-antenna} shows the recognition accuracy across three different datasets, demonstrating that the proposed NN works effectively with various recognition scenarios.
Specifically, the synthesized dataset achieves the highest recognition accuracy, as it contains only $N_{\text{c}}=5$ categories with significant differences between classes.
The sensing performance improves as the number of antennas increases, reaching optimal recognition accuracy at $N_{\text{tr}}=5$ in our simulations. Significant performance gains are observed when the antenna count increases from 1 to 2. However, the improvement becomes progressively smaller as $N_{\text{tr}}$ exceeds 2.
Additionally, increasing $N_{\text{tr}}$ leads to a larger number of trainable parameters in the proposed NN, which results in longer training time.

\begin{figure}[t]
    \centering
    \includegraphics[width=0.78\linewidth]{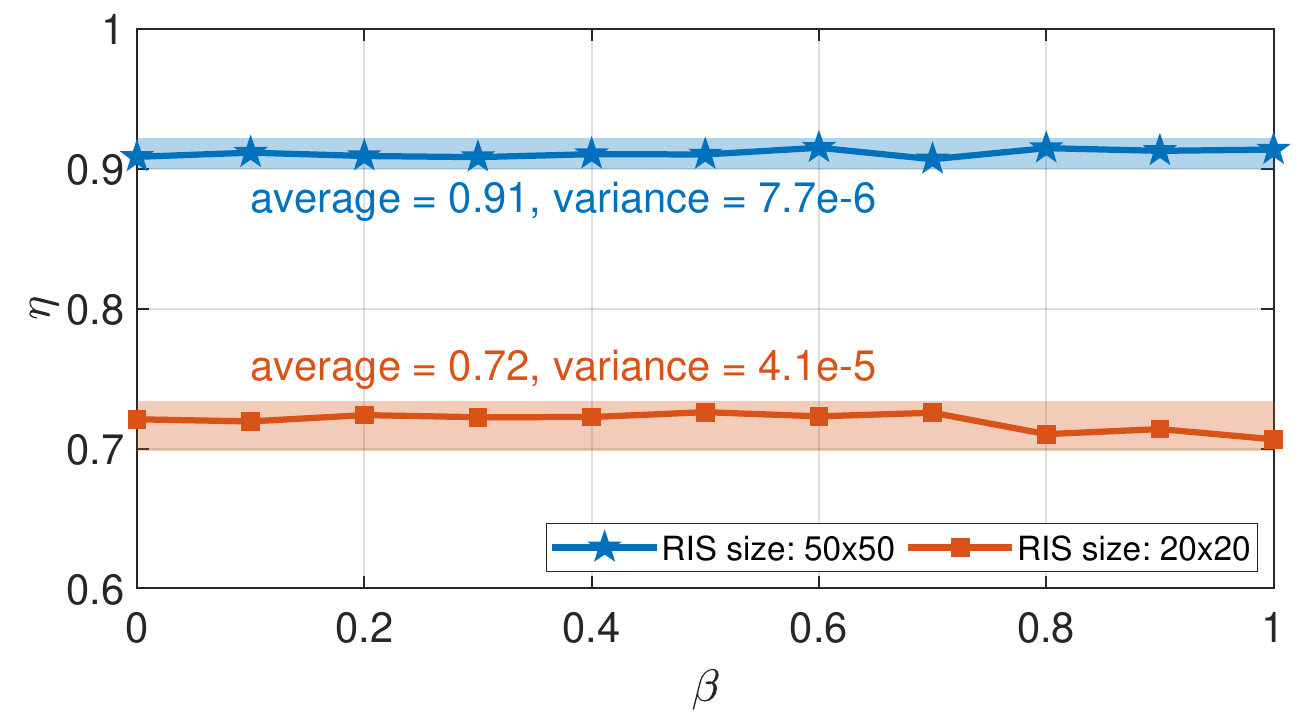}
    \captionsetup{font=footnotesize}
    \caption{Recognition accuracy $\eta$ with various SI levels $\beta$.}
    \label{fig-result-self-interference}
\end{figure}

\item \textbf{SI Levels:}
Finally, we explore the effect of SI on recognition accuracy by varying the SI ratio $\beta$.
The distance from the TX to the RX is $10\lambda_0$, and the antenna radiation patterns are assumed to be isotropic.
The RIS size is either $50\times 50$ or $20\times 20$, with $K=6$ measurements and $N_{\text{tr}}=2$ antennas. Using half of the MNIST dataset for training, the prediction accuracy is shown in Fig. \ref{fig-result-self-interference}.
The results indicate that SI has minimal impact on sensing performance, as the constant residual interference can be learned from the large training dataset.
Therefore, the proposed method demonstrates strong adaptability in full-duplex systems with varying levels of SI, provided that the TX and RX can reliably transmit and receive signals.

\end{itemize}

In summary, the above simulation results validate the effectiveness of the proposed NN across different scenarios involving multiple subcarriers, transceiving antennas, various datasets, and SI levels.
To balance sensing performance and computational complexity in subsequent simulations, we set $N_{\text{f}}=1$, $N_{\text{t}}=N_{\text{r}}=2$, and $\beta=0$ for the following subsections. These settings offer a practical trade-off between efficiency and accuracy, and the insights gained from these configurations can be extended to systems with different setups.
Moreover, half of the MNIST dataset is used for NN training.

\subsubsection{Performance Comparison of Various RIS Phase Designs}
\label{sec-simu-compare-lisp}
We evaluate the correct prediction rates for different RIS phase designs. The LISP \cite{del2020learned} and random \cite{zhao2023intelligent} configurations are used as comparable baselines, both utilizing FC networks with two hidden layers containing 256 neurons.
All methods are subjected to the same training strategy as the proposed NN.
Additionally, we compare the classification accuracy of different RNN structures, specifically the LSTM and GRU units, while keeping the other NN parameters the same.

\begin{figure}[t]
    \centering
    \includegraphics[width=0.78\linewidth]{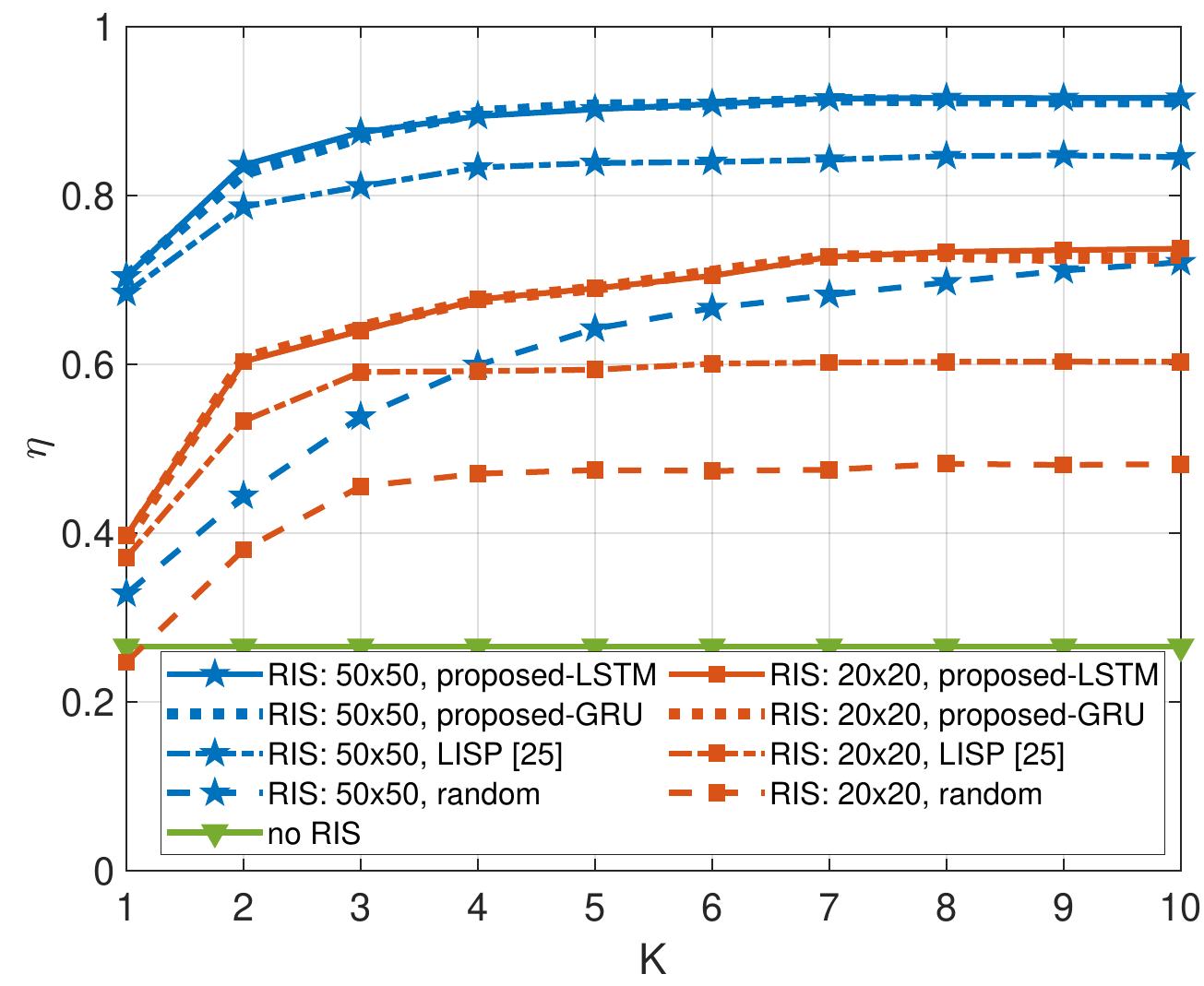}
    \captionsetup{font=footnotesize}
    \caption{Comparison of $\eta$ with different RIS array sizes and configurations.}
    \label{fig-result1-K}
\end{figure}

The simulation results, conducted at a distance of $D = 50\lambda_0$, are illustrated in Fig. \ref{fig-result1-K}.
The results show significant performance improvements of our proposed NNs over both the LISP method and random configurations.
Although the proposed NN has a higher number of trainable parameters compared to the LISP method, the LISP's sensing performance does not improve significantly even with an increased number of neurons \cite{del2020learned}. Notably, the recognition accuracy $\eta$ grows as the number of measurements $K$ increases, reaching a saturation point for our method when $K\ge 7$.
In contrast, the $\eta$ for the LISP method remains relatively unchanged for $K\ge5$.
Moreover, the LSTM and GRU units achieve nearly identical recognition performance.
Enhancing the RIS array size can improve $\eta$; however, our method demonstrates more significant increases in $\eta$ with a smaller RIS.
A $20\times20$ RIS employing $K=10$ of our proposed configurations achieves comparable sensing accuracy to a $50\times50$ RIS with random phase shifts, potentially offering savings on hardware costs.
Comparing scenarios with and without the RIS highlights the benefits of incorporating the RIS to aid in target recognition.

The time required to recognize the target class in this scenario is less than 0.1 seconds, with each frame lasting 10 ms. This suggests that real-time sensing is feasible, even with intermittent sensing intervals as used in the proposed protocol.

\subsubsection{Influence of Distance, Transmit Power, and Training Data Size} 
\label{sec-simu-noise-distance}
\begin{figure}[t]
    \centering
    \includegraphics[width=0.86\linewidth]{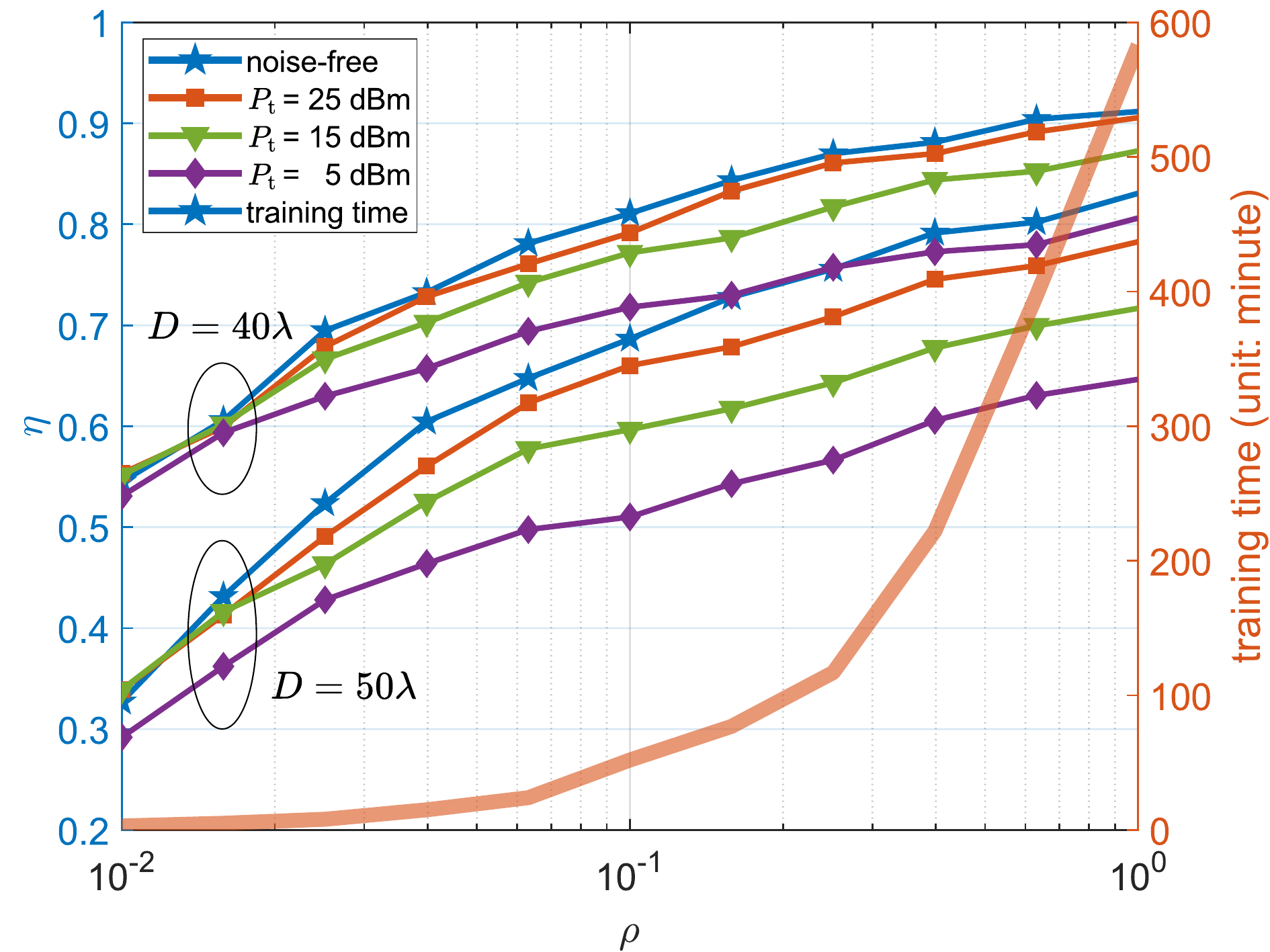}
    \captionsetup{font=footnotesize}
    \caption{Recognition accuracy $\eta$ with respect to the sensing distance $D$, the transmit power $P_{\text{t}}$, and the training dataset ratio $\rho$.}
    \label{fig-result2-distance-noise}
\end{figure}

To assess the impact of target distance $D$, transmit power $P_{\text{t}}$, and training data size, we utilize $K = 3$ measurements for simulations.
The RIS size is set to $40\times 40$.
The additive noise at the RX is randomly generated during both the training and testing phases.
Specifically, we employ the LS channel estimation algorithm to derive $\widehat{\mathbf{H}}_{\text{sen}}$ with the received signals, and the DL signal matrix $\mathbf{X}$ is defined as a discrete Fourier transform matrix for simplicity \cite{lu2024random}.
The simulation results, presented in Fig. \ref{fig-result2-distance-noise}, show how $\rho$ affects the recognition performance, where $\rho$ is the ratio of the number of training samples with respect to the total available $M = 60,000$ instances. Training time was specifically noted for a noise-free environment at the distance of $D=40\lambda_0$.

These results demonstrate that an increase in training data generally leads to higher sensing accuracy, albeit with a growth in training time.
However, once the training phase is completed, the NN can achieve channel estimation, measurement processing, and RIS phase generation in less than 1 ms per instance when using an Nvidia 3090 GPU.
Note that the training time increases linearly with $\rho$, although Fig. \ref{fig-result2-distance-noise} may suggest an exponential relationship due to the use of a logarithmic horizontal axis and a linear vertical axis.
When $P_{\text{t}}=25\ \text{dBm}$, the recognition accuracy $\eta$ approximates that under noise-free scenarios.
However, the sensing performance decreases as $P_{\text{t}}$ is reduced.
Additionally, a shorter distance $D$ significantly enhances recognition accuracy and helps mitigate the adverse effects of additive noise.

\subsubsection{NN Adaptation to Varying Noise Levels}
\label{sec-simu-adaptation}
We explore the adaptability of the proposed NN to varying noise levels, assuming a constant transmit signal power of $P_{\text{t}} = 25\ \text{dBm}$.
Other parameters remain the same as those in Sec. \ref{sec-simu-noise-distance}. The simulation results are summarized in Table \ref{tab-adaptation}.
In ``Scenario 1,'' the noise power used during both training and testing is the same, leading to high recognition accuracy. The average training time for the NN is around 210 minutes.
Alternatively, in ``Scenario 2'' we test the NN---originally trained with no additive noise---on samples with varying noise levels. The recognition accuracy for these tests shows that minor noise results in a slight performance drop compared to ``Scenario 1,'' but significant noise levels, especially with noise power at -60 dBm, lead to considerable degradation in sensing performance.

To mitigate this, the NN's classifier can be retrained with noisy measurements, which takes less than 2 minutes---less than 1\% of the time required to train the NN from scratch. The classification accuracy after retraining is shown as ``Scenario 3'' in Table \ref{tab-adaptation}, and is comparable to ``Scenario 1'' when the noise power is -70 or -80 dBm. This indicates that the proposed NN can quickly adapt to minor noise by retraining the classifier.
However, when noise levels are significantly high (e.g., -60 dBm), the retrained NN cannot fully recover the accuracy of ``Scenario 1,'' though there is an improvement over ``Scenario 2''. This suggests that the radio environment may have changed substantially, making the RIS configurations optimized for noise-free conditions less effective. In such cases, it is recommended to thoroughly retrain the NN to restore high recognition accuracy.

\begin{table}[t]
    \renewcommand{\arraystretch}{1.4}
    \centering
    \fontsize{8}{8}\selectfont
    \captionsetup{font=small}
    \captionof{table}{Adaptation of the proposed NN  to varying noise levels by retraining the classifier.}\label{tab-adaptation}
    \begin{threeparttable}
        \begin{tabular}{ccccc}
            \specialrule{1pt}{0pt}{-1pt}\xrowht{10pt}
            Noise power (unit: dBm) & -60 & -70 & -80 & No noise \\
            \hline
            Scenario 1 & 0.8053 & 0.8634 & 0.8957 & 0.9000 \\
            Scenario 2 & 0.3583 & 0.7724 & 0.8905 & / \\
            Scenario 3 & 0.6760 & 0.8432 & 0.8921 & / \\
            \specialrule{1pt}{0pt}{0pt}
        \end{tabular}
    \end{threeparttable}
\end{table}

\begin{figure}[t]
    \centering
    \includegraphics[width=0.78\linewidth]{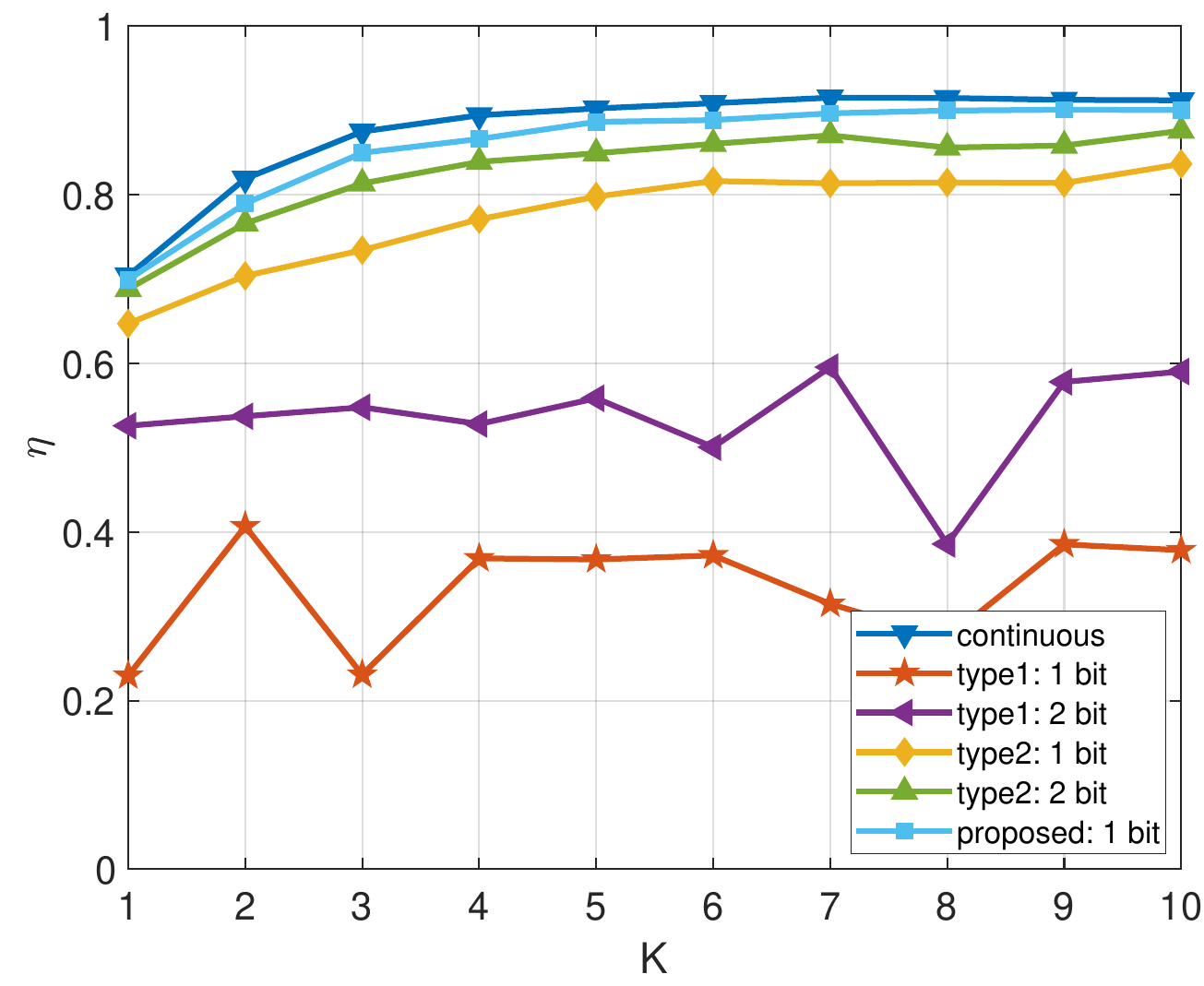}
    \captionsetup{font=footnotesize}
    \caption{Sensing performance comparison when implementing discrete RIS phases with different methods.}
    \label{fig-result-discrete}
\end{figure}

\begin{table}[t]
    \renewcommand{\arraystretch}{1.4}
    \centering
    \fontsize{8}{8}\selectfont
    \captionsetup{font=small}
    \captionof{table}{Sensing performance loss of $\eta$ by directly quantifying the optimized continuous RIS phases (``type1'' in Fig. \ref{fig-result-discrete}) when $K=5$.}\label{tab-discrete}
    \begin{threeparttable}
        \begin{tabular}{cccccccc}
            \specialrule{1pt}{0pt}{-1pt}\xrowht{10pt}
            Quantization bit & 1 & 2 & 3 & 4 & 5 & 6 & 7 \\
            \hline
            Performance loss of $\eta$ & 59\% & 38\% & 21\% & 6\% & 3\% & 1\% & 0 \\
            \specialrule{1pt}{0pt}{0pt}
        \end{tabular}
    \end{threeparttable}
\end{table}

\subsubsection{Performance Evaluation with Discrete RIS Phases}
\label{sec-simu-discrete}
The influences of quantified RIS phase shifts are discussed in Fig. \ref{fig-result-discrete} and Table \ref{tab-discrete} by employing a $50\times50$ RIS.
Specifically, ``type1'' in Fig. \ref{fig-result-discrete} denotes that the NN is trained without considering the quantization constraints, and the RIS configurations are obtained from the direct discretization of optimized continuous values.
The discretized RIS phases lead to a mismatch between the acquired CSI measurements and the classifier parameters, significantly degrading the sensing performance when the quantization bit is one or two.
Table \ref{tab-discrete} further presents the performance loss of ``type1'' with different discretization bits compared to continuous RIS phases, implying that the ``type1'' method works well when five or higher quantization bits are available.

Alternatively, ``type2'' in Fig. \ref{fig-result-discrete} further retrains the classifier in the proposed NN with measurements derived from discretized RIS phases, a method similar to ``Scenario 2'' in Table \ref{tab-adaptation}.
Hence, the mismatch in ``type1'' can be partly released, and the classification accuracy is enhanced considerably.
However, the discretized RIS phases are not the sub-optimal values directly optimized by the NN, potentially degrading the sensing performance.

Finally, we exploit the proposed method in Sec. \ref{sec-discrete-ris-phase} and train the NN until the values in $\mathbf{v}_{n_{\text{s}}}'$ of \eqref{eq-vns} possess a discrepancy less than 0.0001 with 0 or 1.
Hence, the quantization constraints are fully utilized during NN training, and the generated RIS phases are ensured to be discrete values.
The hyperparameter $\phi$ is selected by training a series of models with sweeping values in the range $[0.001, 0.5]$, with the best-performing model on the validation dataset being chosen. According to the simulation results in Fig. \ref{fig-result-discrete}, sensing performance comparable to that of continuous RIS phases can be achieved even with 1-bit quantized RIS phases.

\begin{table}[t]
    \renewcommand{\arraystretch}{1.4}
    \centering
    \fontsize{8}{8}\selectfont
    \captionsetup{font=small}
    \captionof{table}{Performance comparison between ``recognition'' and ``imaging-recognition''.}\label{tab-first-imaging}
    \begin{threeparttable}
        \begin{tabular}{cc p{0.8cm}<{\centering}}
            \specialrule{1pt}{0pt}{-1pt}\xrowht{10pt}
            Method & Measurement No. ($K$) & $\eta$ \\
            \hline
            \multirow{5}{*}{Imaging-Recognition} & 10 & 0.5521 \\
            & 20 & 0.6604 \\
            & 30 & 0.7077 \\
            & 40 & 0.7734 \\
            & 50 & \textbf{0.8011} \\
            \hline
            Recognition: LISP \cite{del2020learned} & 10 & 0.7008 \\
            Recognition: Proposed & 10 & \textbf{0.8020} \\
            \specialrule{1pt}{0pt}{0pt}
        \end{tabular}
    \end{threeparttable}
\end{table}

\subsubsection{Performance Comparison with ``Imaging-Recognition''}
\label{sec-simu-first-imaging}
We compare the target recognition performances between the direct classification of CSI measurements and the classification of formulated radio images from CSI measurements.
We employ a CNN for radio image classification, where the CNN includes two convolution layers with 16 and 32 channels, respectively, both followed by a maximum pooling layer.
The output of the second convolution layer is flattened and input to an FC layer, calculating the classification probabilities.

To obtain the images of the targets, we consider the single-view imaging scenario in our previous work \cite{huang2024ris}, harnessing the compressed sensing-based generalized approximate message passing (GAMP) algorithm and optimizing the RIS phase shifts for imaging.
However, only the TX-ROI-RIS-RX path is employed for imaging in \cite{huang2024ris}, which requires subpath extraction from raw CSI measurements.
To simplify the simulations, we modify the physical model in the proposed NN and only keep the TX-ROI-RIS-RX path in the CSI measurements, releasing the subpath extraction process.

The sensing performances of ``recognition'' and ``imaging-recognition'' are compared in Table \ref{tab-first-imaging}, showing a remarkable performance gain of the LISP \cite{del2020learned} and the proposed method over the ``imaging-recognition'' method.
Due to the severe lack of CSI measurements, the generated images are seriously distorted, and nearly no visual information can be acquired by manually observing the images.
Consequently, only the recognition accuracy of 55.21\% is achieved with 10 measurements.
Considering the errors introduced in the subpath extraction process, the sensing performance may be even lower.
Thanks to the powerful classification ability of the CNN, triple and quintuple measurements are desired to achieve comparable performances to the LISP \cite{del2020learned} and the proposed method.
Therefore, the proposed task-oriented recognition scheme successfully reduces the measurement overhead and degrades the communication resource occupation.

\begin{table}[t]
    \renewcommand{\arraystretch}{1.4}
    \centering
    \fontsize{8}{8}\selectfont
    \captionsetup{font=small}
    \captionof{table}{SE comparison with the LOS path (unit: bit/s/Hz).}\label{tab-se-1}
    \begin{threeparttable}
        \begin{tabular}{ccccc}
            \specialrule{1pt}{0pt}{-1pt}\xrowht{10pt}
            RIS size & ${\text{SE}}(\boldsymbol{\omega}_{\text{com}})$ & ${\text{SE}}(\boldsymbol{\omega}_{\text{sen}})$ & $\overline{\text{SE}}_{\mu=1}$ & SE Loss \\
            \hline
            $10\times10$ & 50.40 & 58.14 & 58.09 & 0.10\% \\
            $20\times20$ & 51.35 & 69.44 & 69.31 & 0.19\% \\
            $30\times30$ & 50.97 & 78.94 & 78.74 & 0.25\% \\
            \specialrule{1pt}{0pt}{0pt}
        \end{tabular}
    \end{threeparttable}
\end{table}

\begin{table}[t]
    \renewcommand{\arraystretch}{1.4}
    \centering
    \fontsize{8}{8}\selectfont
    \captionsetup{font=small}
    \captionof{table}{SE comparison without the LOS path (unit: bit/s/Hz).}\label{tab-se-2}
    \begin{threeparttable}
        \begin{tabular}{ccccc}
            \specialrule{1pt}{0pt}{-1pt}\xrowht{10pt}
            RIS size & ${\text{SE}}(\boldsymbol{\omega}_{\text{com}})$ & ${\text{SE}}(\boldsymbol{\omega}_{\text{sen}})$ & $\overline{\text{SE}}_{\mu=1}$ & SE Loss \\
            \hline
            $10\times10$ & 23.00 & 46.72 & 46.55 & 0.36\% \\
            $20\times20$ & 24.86 & 65.51 & 65.22 & 0.44\% \\
            $30\times30$ & 27.30 & 76.79 & 76.43 & 0.46\% \\
            \specialrule{1pt}{0pt}{0pt}
        \end{tabular}
    \end{threeparttable}
\end{table}

\begin{figure}[t]
    \centering
    \includegraphics[width=0.86\linewidth]{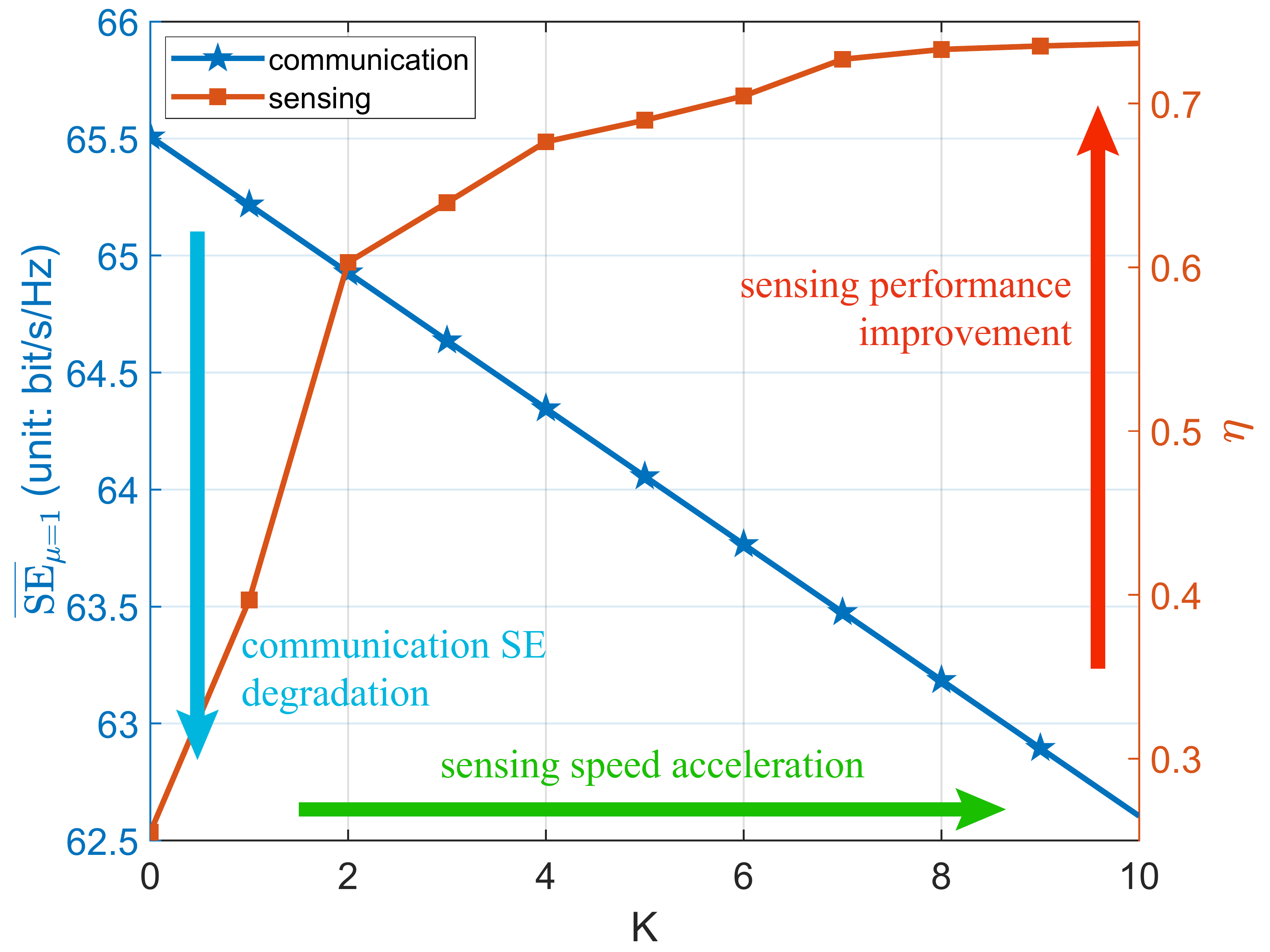}
    \captionsetup{font=footnotesize}
    \caption{Tradeoff between communication and sensing performance by varying the number of measurements in one frame.}
    \label{fig-tradeoff}
\end{figure}

\subsubsection{Communication Performance Analysis}
\label{sec-simu-communication}
The SE performance, based on the proposed RIS configurations and protocol, is enumerated in Tables \ref{tab-se-1} and \ref{tab-se-2}.
The analysis considers ${\mu=1}$, ${P_{\text{t}}=12\ \text{dBm}}$, $B = 20\ \text{MHz}$, $N_{\text{f}}=5$, and the presence or absence of the LOS path between the UE and the BS.
The SE loss incurred to facilitate target recognition is calculated as $[{\text{SE}}(\boldsymbol{\omega}_{\text{com}}) - \overline{\text{SE}}_{\mu=1}]/{\text{SE}}(\boldsymbol{\omega}_{\text{com}})$.
According to \cite{wu2018intelligent}, the optimal RIS phase configuration $\boldsymbol{\omega}_{\text{com}}$ depends on the radio environment and varies with the target in the ROI. Therefore, the simulation results are presented using the first sample from the training dataset as a representative example.

The results show that the SE is generally higher with an LOS path compared to when the LOS path is blocked.
When RIS phases are optimized to enhance classification accuracy, ${\text{SE}}(\boldsymbol{\omega}_{\text{sen}})$ experiences a reduction compared to ${\text{SE}}(\boldsymbol{\omega}_{\text{com}})$, especially with the absence of the LOS path.
Nevertheless, the average SE shows only a marginal loss compared to ${\text{SE}}(\boldsymbol{\omega}_{\text{com}})$, typically lower than 0.5\% in the scenarios considered. This loss can be further minimized when the numerology $\mu > 1$, as the phase shift $\boldsymbol{\omega}_{\text{sen}}$ is applied for only $N_{\text{t}}$ symbol intervals, which is far fewer than the total number of symbols $N_0$ in one frame. Therefore, the proposed approach has minimal impact on communication performance.

To better illustrate the tradeoff between communication and sensing performance, we vary the number of measurements $K$ in one frame. The simulation results with a $20 \times 20$ RIS and no LOS path are shown in Fig. \ref{fig-tradeoff}, reusing the data from Fig. \ref{fig-result1-K}. A larger $K$ results in higher consumption of communication resources, as more symbol intervals are allocated for sensing. However, capturing $K = 10$ measurements only utilizes 20 symbol intervals, which is small compared to $N_0 = 280$. Consequently, sensing performance can be significantly improved with only a minor reduction in communication SE.

\begin{figure*}[t]
    \centering
    \includegraphics[width=0.8\linewidth]{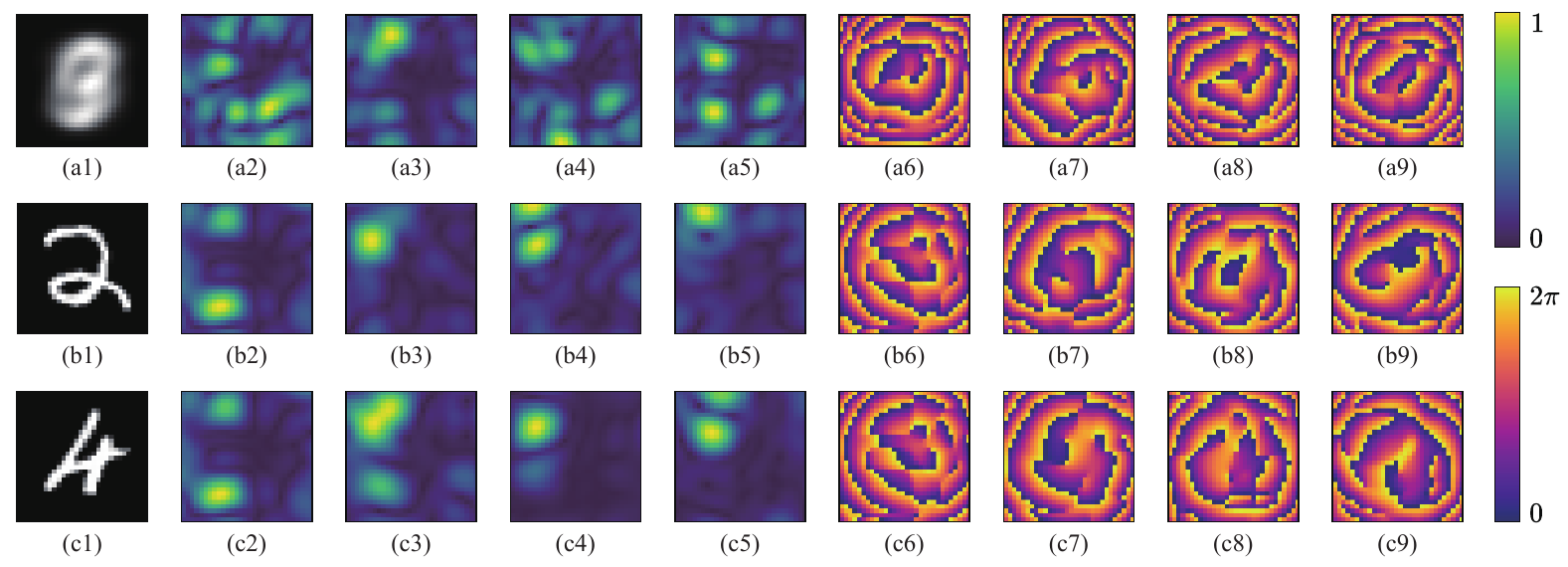}
    \captionsetup{font=footnotesize}
    \caption{Visualization of the generated RIS phases of the proposed method and the LISP method \cite{del2020learned} when $K=4$. The first row draws the results of the LISP method; the second and third rows depict the results of the proposed method for two different targets, respectively. The first column presents the target being sensed; the second to fifth and sixth to ninth columns are the amplitudes and phases of the RIS radiation patterns, respectively.}
    \label{fig-result-visual}
\end{figure*}

\subsubsection{Interpretation of the Optimized RIS phases}
\label{sec-simu-interpretation}
First, we visualize the RIS radiation patterns generated by the proposed NN and the LISP method \cite{del2020learned} with $N_{\text{s}} = 20\times 20$.
However, the considered channel model involves multiple paths, making the RIS radiation patterns hard to be presented.
Hence, we take a similar modification to Sec. \ref{sec-simu-first-imaging}, reserving only the TX-RIS-ROI-RX path and a single TX antenna.
Hence, the RIS radiation pattern at the $k$-th moment can be calculated by $\boldsymbol{\varsigma}_k = \mathbf{H}_{\text{ris-roi}} \text{diag}(\boldsymbol{\omega}_k) \mathbf{h}_{\text{tx-ris}}\in\mathbb{C}^{N_{\text{i}}\times 1}$.
Reshaping $\boldsymbol{\varsigma}_k$ into a $30\times30$ matrix, its visualization can be found in Fig. \ref{fig-result-visual} with $K=4$, where the first row represents the LISP method, while the second and third rows denote the results of the proposed method.
The first column of Fig. \ref{fig-result-visual} pictures the target being sensed, the second to the fifth columns draw the absolute values of $\boldsymbol{\varsigma}_k$, and the sixth to the ninth columns depict the phases of $\boldsymbol{\varsigma}_k$.
In Fig. \ref{fig-result-visual}(a1), we present the average of the 10,000 test images, finding that the targets overlap around the ROI center.
Fig. \ref{fig-result-visual}(a2) to Fig. \ref{fig-result-visual}(a5) imply that the LISP method tends to perceive the surrounding regions of the ROI, where the targets may exhibit distinctive features.
Different from the LISP method that illuminates the RIS beams to wide ranges of the ROI, the proposed method tends to generate RIS patterns focused on certain areas.
Moreover, the proposed method customizes RIS phases for distinct targets, as illustrated in the second and third rows in Fig. \ref{fig-result-visual}.

\begin{figure}[t]
    \centering
    \includegraphics[width=0.9\linewidth]{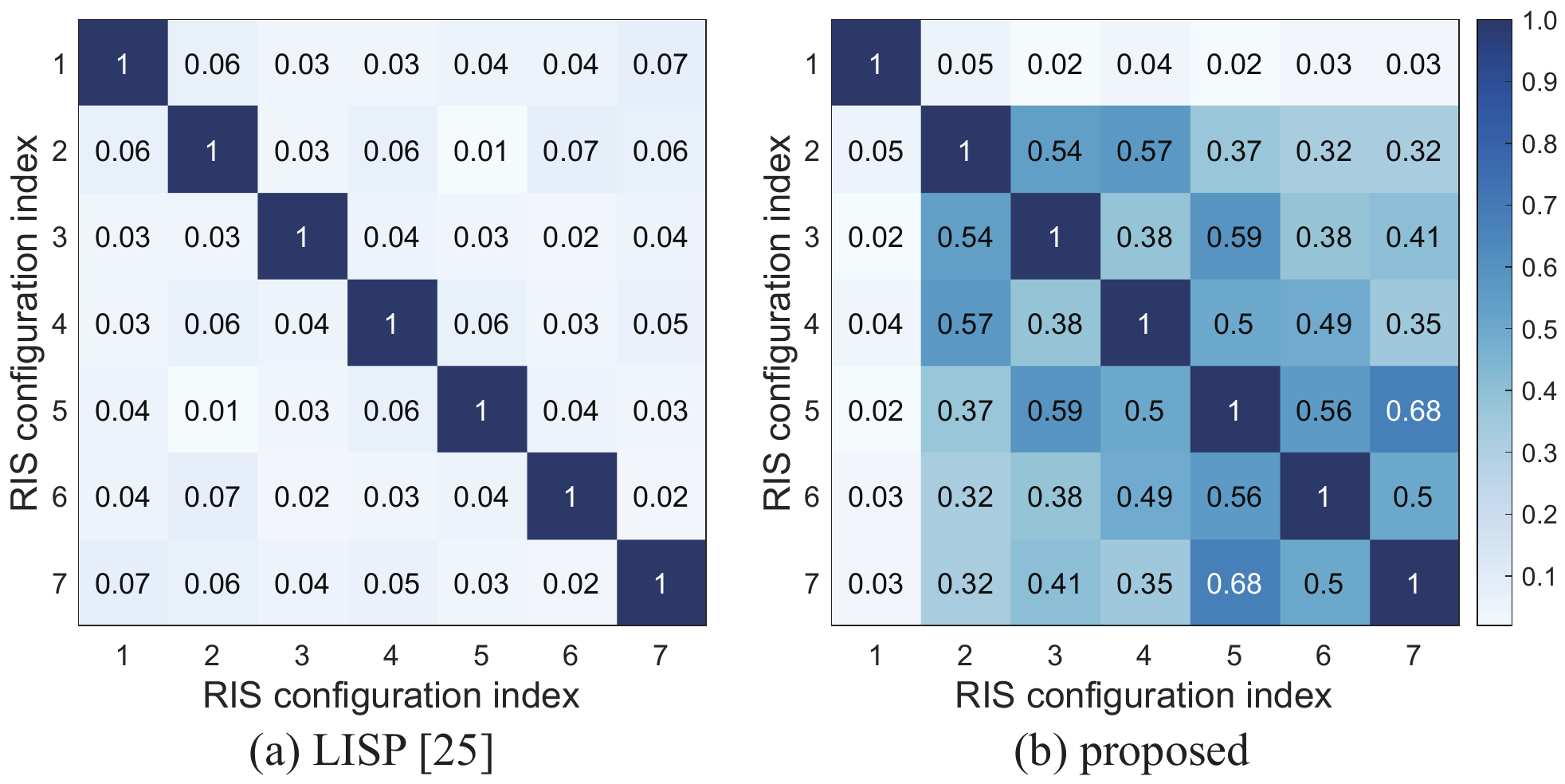}
    \captionsetup{font=footnotesize}
    \caption{Comparison of the correlations of the RIS phase configurations.}
    \label{fig-result3-heat}
\end{figure}

Second, we compare the correlation patterns of the RIS phase shifts produced by these two methods.
This comparison is depicted in Fig. \ref{fig-result3-heat} for $K=7$, where the RIS phase correlations are computed using the formula $|\boldsymbol{\omega}^{\text{H}}_{k_1}\boldsymbol{\omega}_{k_2}| / \|\boldsymbol{\omega}_{k_2}\|_2\|\boldsymbol{\omega}_{k_2}\|_2$, where $k_1, k_2 = 1, 2, \ldots, K$.
In Fig. \ref{fig-result3-heat}(a), it is observed that the RIS configurations produced by the LISP method exhibit minimal correlations, thereby capturing pseudo-orthogonal information relevant to any target class.
Conversely, as illustrated in Fig. \ref{fig-result3-heat}(b), the RIS patterns generated by the proposed NN show relatively high correlations from $k=2$ onwards.
Given that the RIS phase shifts are uniquely tailored for each target, these results are averaged over 10,000 test data samples.
Given our objective to capture the most pertinent information for classifying the target, where each target's category remains constant, it is reasonable to produce correlated RIS configurations.
Thus, although the proposed method may not gather as much information as the LISP method, the information it does capture is specifically optimized for the targets, making it exceedingly valuable for the final classification task.
The above phenomena conclude that the proposed method works significantly different from the LISP method \cite{del2020learned}.

\begin{table}[t]
    \renewcommand{\arraystretch}{1.4}
    \centering
    \fontsize{8}{8}\selectfont
    \captionsetup{font=small}
    \captionof{table}{Performance evaluation of the learned decisioner.}\label{tab-decision}
    \begin{threeparttable}
        \begin{tabular}{c p{1.0cm}<{\centering} p{1.0cm}<{\centering}}
            \specialrule{1pt}{0pt}{-1pt}
            \xrowht{10pt} Method & $\overline{K}$ & $\eta$ \\
            \hline
            \rowcolor{gray!20} Model-A-1 ($K_{\text{m}}=2$) & 2 & 0.6030 \\
            Model-A-2 ($K_{\text{m}}=3$) & 3 & 0.6395 \\
            \rowcolor{gray!20} Model-A-3 ($K_{\text{m}}=4$) & 4 & 0.6766 \\
            Model-A-4 ($K_{\text{m}}=5$) & 5 & 0.6898 \\
            \rowcolor{gray!20} Model-A-5 ($K_{\text{m}}=6$) & 6 & 0.7046 \\
            \hline
            Model-B-1 ($K_{\text{m}}=6,\gamma=1.14$) & 2.947 & 0.6641 \\
            \rowcolor{gray!20} Model-B-2 ($K_{\text{m}}=6,\gamma=1.08$) & 3.450 & 0.6779 \\
            Model-B-3 ($K_{\text{m}}=8,\gamma=1.10$) & 4.079 & 0.6888 \\
            \rowcolor{gray!20} Model-B-4 ($K_{\text{m}}=8,\gamma=1.06$) & 4.556 & 0.6961 \\
            \hline
            Model-C-1 ($K_{\text{m}}=6,\zeta=0.50$) & 2.800 & 0.6026 \\
            \rowcolor{gray!20} Model-C-2 ($K_{\text{m}}=6,\zeta=0.60$) & 3.560 & 0.6379 \\
            Model-C-3 ($K_{\text{m}}=6,\zeta=0.70$) & 4.229 & 0.6555 \\
            \rowcolor{gray!20} Model-C-4 ($K_{\text{m}}=6,\zeta=0.80$) & 4.778 & 0.6629 \\
            \specialrule{1pt}{0pt}{0pt}
        \end{tabular}
    \end{threeparttable}
\end{table}

\subsubsection{Performance Evaluation of the Learned Decisioner} 
\label{sec-simu-decisioner}
We evaluate the performance of the enhanced recognizer proposed in Sec. \ref{sec-decision} with dynamic sensing durations.
We employ a $20\times20$ RIS, and the simulation results are shown in Table \ref{tab-decision}, where $\overline{K}$ denotes the average measurement number over the 10,000 testing samples.
In the table, `Model-A'' refers to the proposed model without the decisioner from Sec. \ref{sec-ris-phase-design}; ``Model-B'' refers to the proposed model with the learned decisioner from Sec. \ref{sec-decision}; and ``Model-C'' refers to the model with a threshold-based decisioner \cite{wang2020glance}, which decides when to terminate the sensing process by comparing the maximum classification probability in $\mathbf{p}_k$ to a given threshold $\zeta$.

The results in Table \ref{tab-decision} demonstrate the superiority of the learned decisioner.
For example, Model-A-2 uses 3 measurements to achieve a recognition accuracy ($\eta$) of 63.95\%. However, Model-B-1, with $\gamma=1.14$, achieves a classification accuracy of 66.41\% using only 2.947 measurements, approaching the performance of Model-A-3, which requires 4 measurements. The performance improvement in Model-B-1 comes from the learned policy that dynamically terminates the sensing process, allocating a customized number of measurements for each target based on its classification difficulty.
In contrast, Model-C-2, with a threshold $\zeta=0.60$, only achieves an $\eta$ of 63.79\% using 3.56 measurements, performing worse than Model-A-2. This illustrates that a fixed termination policy based on a constant threshold is less effective than a policy learned through training.
Other comparisons in Table \ref{tab-decision} yield similar conclusions, further validating the effectiveness of the proposed Model-B with a learned decisioner.

\begin{figure}[t]
    \centering
    \includegraphics[width=0.78\linewidth]{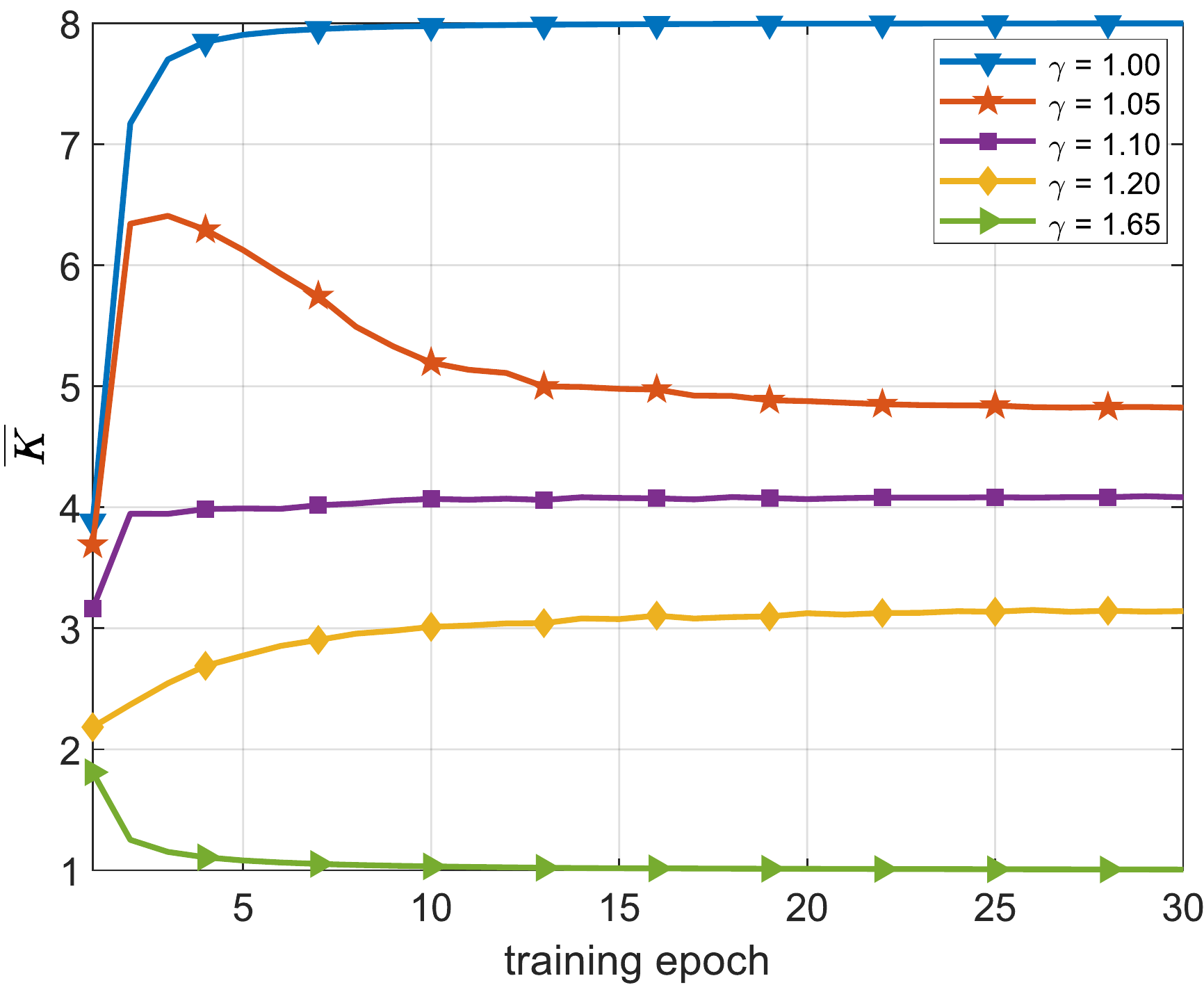}
    \captionsetup{font=footnotesize}
    \caption{Average measurement number $\overline{K}$ versus NN training epochs.}
    \label{fig-result-penalty-1}
\end{figure}

\subsubsection{Influences of the Penalty Factor on the Learned Decisioner}
\label{sec-simu-penalty}

We discuss the influences of the penalty factor $\gamma$ on the proposed decisioner and depict the changes of the average measurement number $\overline{K}$ with various $\gamma$ during the training phase in Fig. \ref{fig-result-penalty-1}, where $K_{\text{m}}=8$.
The NN converges to a state that always ends with $k=8$ measurements when $\gamma=1$, since the NN is not forced to exit the sensing process earlier and explores to obtain the lowest cross-entropy loss, where more CSI measurements typically obtain richer information about the target.
We can find that $\overline{K}$ decreases with $\gamma$ increasing, since the loss function penalizes the NN more when it terminates with larger measurement numbers.
Interestingly, when $\gamma=1.05$, $\overline{K}$ first increases to employ large numbers of measurements for target recognition and subsequently decreases to strike a balance between the measurement number and sensing accuracy.
$\overline{K}$ monotonically increases when $\gamma=1.1$ or $\gamma=1.2$ during NN training, since the degraded cross-entropy loss with more measurements surpasses the corresponding penalty.
When $\gamma$ is extremely large, e.g., $\gamma=1.65$ in the considered scenario, the NN is trained to terminate the sensing procedure with only one measurement.

\begin{figure}[t]
    \centering
    \includegraphics[width=0.78\linewidth]{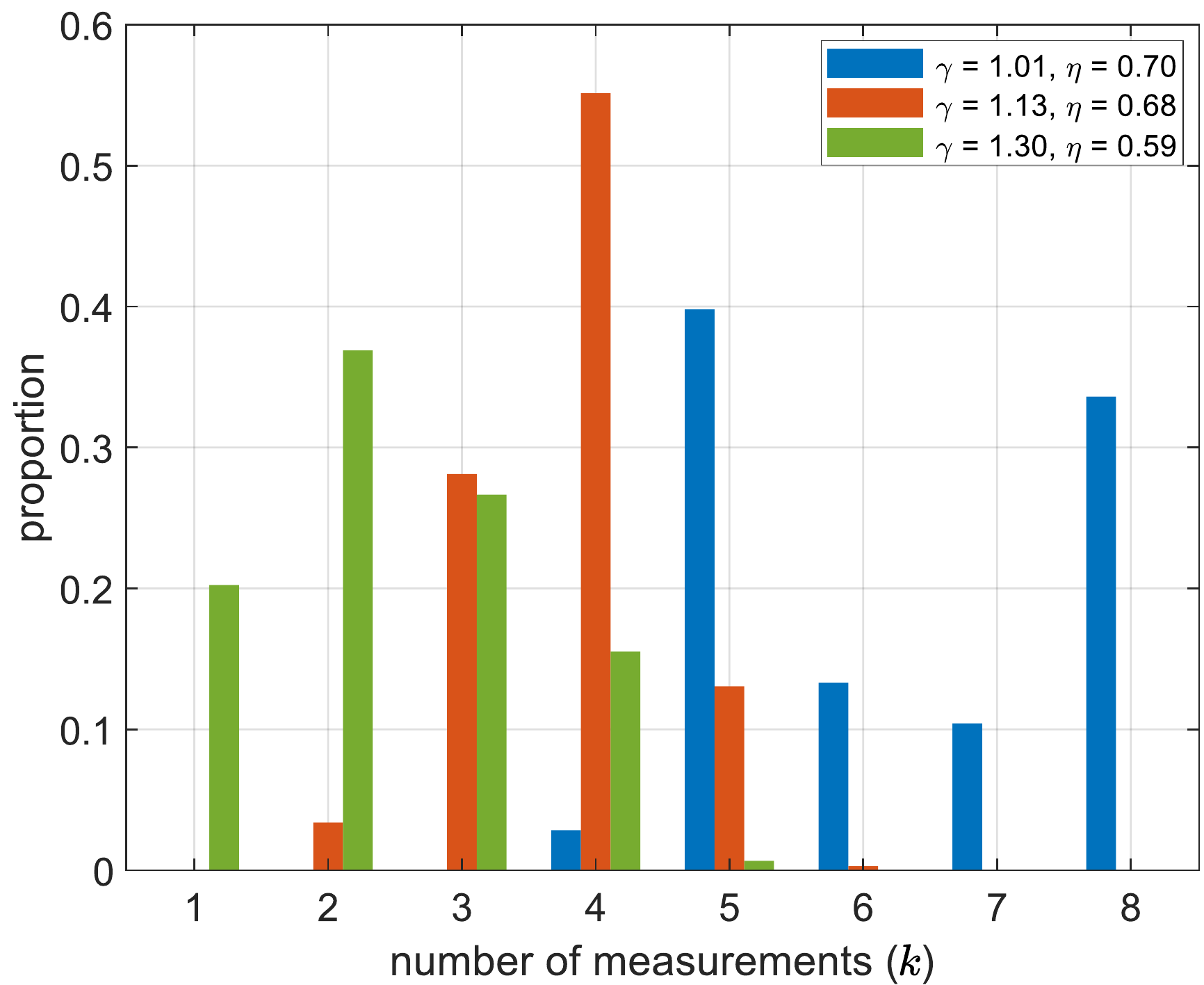}
    \captionsetup{font=footnotesize}
    \caption{Distributions of the measurement numbers with the learned decisioner.}
    \label{fig-result-penalty-2}
\end{figure}

Furthermore, we depict the distributions of the measurement numbers with the learned decisioner in Fig. \ref{fig-result-penalty-2}.
The NN with $\gamma=1.01$ tends to stop sensing when $k=5$ or $k=8$, preliminarily revealing the impact of $\gamma$ when compared to the NN with $\gamma=1$ that always ends at $k=8$.
Increasing $\gamma$ to 1.13 guides the NN to terminate the sensing procedure at $k=4$ for most testing samples.
The NN with $\gamma=1.3$ further decreases the average measurement number, preferring to end at the first three steps.
Thus, the penalty factor $\gamma$ can effectively tune the termination preference of the decisioner.

\section{Conclusion}

This study presents a RIS-aided communication system where the DL signals are multiplexed to recognize the target in the ROI. A time-division protocol is proposed to configure the RIS phase shifts according to different purposes. To achieve high classification accuracy, we propose a novel NN architecture, which cooperates with the physical channel model and fully harnesses prior information about the scene, task, target, and RIS phase quantization. The RIS configurations are tailored for each target and trained together with the NN parameters. Furthermore, we introduce a decisioner to dynamically terminate the sensing procedure according to the acquired information, maximizing the efficiency of resource utilization. Therefore, both the RIS configurations and sensing durations are customized for the targets being sensed. Simulation results demonstrate that the proposed algorithm outperforms the state-of-the-art methods, achieving high sensing accuracy with minimal impact on communication performance. Employing the learned decisioner, the number of measurements may be reduced while ensuring certain recognition accuracy.

\begin{appendices}

\section{}\label{appendix-channel}

First, we denote the LOS channel from the TX to the UE in \eqref{eq-commun-channel} as $\mathbf{h}_{\text{tx}\text{-}\text{ue}, n_{\text{f}}} = [h_{\text{tx}\text{-}\text{ue}, n_{\text{f}}, 1}, h_{\text{tx}\text{-}\text{ue}, n_{\text{f}}, 2}, \ldots, h_{\text{tx}\text{-}\text{ue}, n_{\text{f}}, N_{\text{t}}}]^{\text{T}}$.
The $n_{\text{t}}$-th element in $\mathbf{h}_{\text{tx}\text{-}\text{ue}, n_{\text{f}}}$ is given by \cite{goldsmith2005wireless}
\begin{equation}\label{eq-appendix-a-1}
h_{\text{tx}\text{-}\text{ue}, n_{\text{f}}, n_{\text{t}}} = \frac{\lambda_0 \sqrt{G_{\text{com}}}}{{4\pi}d_{n_{\text{t}}, {\text{ue}}}} e^{-j2\pi \frac{d_{n_{\text{t}}, {\text{ue}}}}{\lambda_{n_{\text{f}}}}},
\end{equation}
where $d_{n_{\text{t}}, {\text{ue}}}$ denotes the distance from the $n_{\text{t}}$-th TX antenna to the UE, and $\lambda_{n_{\text{f}}}$ represents the wavelength of the ${n_{\text{f}}}$-th subcarrier.
The combined antenna gain is $G_{\text{com}}=G_{\text{t}}G_{\text{ue}}$, where $G_{\text{t}}$ and $G_{\text{ue}}$ are the antenna gains of the TX and the UE, respectively.

Additionally, we denote $\bar{\boldsymbol{\omega}} = G_{\text{ris}}\boldsymbol{\omega}$, where $G_{\text{ris}} = \sqrt{4\pi}AS/\lambda_0$ is the scattering gain of the RIS element \cite{huang2023joint,tang2020wireless}, with $S$ as the RIS element area and $A$ as the scattering coefficient (set to 1 in this study).
Consequently, the channel expressions are
\begin{equation}\label{eq-000}
\begin{aligned}
&\mathbf{h}_{\text{tx}\text{-}\text{roi}\text{-}\text{ue}, n_{\text{f}}} \!\!\!&= &\ \mathbf{H}_{\text{tx}\text{-}\text{roi}, n_{\text{f}}} \text{diag}(\boldsymbol{\sigma}) \mathbf{h}_{\text{roi}\text{-}\text{ue}, n_{\text{f}}}, \\
&\mathbf{h}_{\text{tx}\text{-}\text{ris}\text{-}\text{ue}, n_{\text{f}}} \!\!\!&= &\ \mathbf{H}_{\text{tx}\text{-}\text{ris}, n_{\text{f}}} \text{diag}(\bar{\boldsymbol{\omega}}) \mathbf{h}_{\text{ris}\text{-}\text{ue}, n_{\text{f}}}, \\
&\mathbf{h}_{\text{tx}\text{-}\text{roi}\text{-}\text{ris}\text{-}\text{ue}, n_{\text{f}}} \!\!\!&= &\ \mathbf{H}_{\text{tx}\text{-}\text{roi}, n_{\text{f}}} \text{diag}(\boldsymbol{\sigma}) \mathbf{H}_{\text{roi}\text{-}\text{ris}, n_{\text{f}}} \text{diag}(\bar{\boldsymbol{\omega}}) \mathbf{h}_{\text{ris}\text{-}\text{ue}, n_{\text{f}}}, \\
&\mathbf{h}_{\text{tx}\text{-}\text{ris}\text{-}\text{roi}\text{-}\text{ue}, n_{\text{f}}} \!\!\!&= &\ \mathbf{H}_{\text{tx}\text{-}\text{ris}, n_{\text{f}}} \text{diag}(\bar{\boldsymbol{\omega}}) \mathbf{H}_{\text{ris}\text{-}\text{roi}, n_{\text{f}}} \text{diag}(\boldsymbol{\sigma}) \mathbf{h}_{\text{roi}\text{-}\text{ue}, n_{\text{f}}}.
\end{aligned}
\end{equation}
Here, $\mathbf{H}_{\text{tx}\text{-}\text{roi}, n_{\text{f}}}\in\mathbb{C}^{N_{\text{t}}\times N_{\text{i}}}$ denotes the channel from the TX to the ROI, with the $(n_{\text{t}}, n_{\text{i}})$-th element given by
\begin{equation}\label{eq-aaa}
h_{\text{tx}\text{-}\text{roi}, n_{\text{f}}, n_{\text{t}}, n_{\text{i}}} = \sqrt{G_{\text{t}}} \times \frac{1}{\sqrt{4\pi} d_{n_{\text{t}}, n_{\text{i}}}}e^{-j2\pi \frac{d_{n_{\text{t}}, n_{\text{i}}}}{\lambda_{n_{\text{f}}}}},
\end{equation}
where $d_{n_{\text{t}}, n_{\text{i}}}$ is the distance between the $n_{\text{t}}$-th TX antenna and the $n_{\text{i}}$-th voxel.
The term $\sqrt{G_{\text{t}}}$ denotes the path gain related to the TX antenna. 
$\mathbf{h}_{\text{roi}\text{-}\text{ue}, n_{\text{f}}}\in\mathbb{C}^{N_{\text{i}}}$ represents the channel from the ROI to the UE. Its $n_{\text{i}}$-th element is expressed as
\begin{equation}\label{eq-bbb}
h_{\text{roi}\text{-}\text{ue}, n_{\text{f}}, n_{\text{i}}} = \lambda_0\sqrt{\frac{G_{\text{ue}}}{4\pi}} \times \frac{1}{\sqrt{4\pi} d_{n_{\text{i}}, \text{ue}}}e^{-j2\pi \frac{d_{n_{\text{i}}, \text{ue}}}{\lambda_{n_{\text{f}}}}},
\end{equation}
where $d_{n_{\text{i}}, \text{ue}}$ is the transmission distance between the $n_{\text{i}}$-th voxel to the UE.
The term $\lambda_0\sqrt{{G_{\text{ue}}}/{4\pi}}$ represents the path gain contributed by the effective aperture of the receiving antenna \cite{tang2020wireless,dardari2020communicating}.
$\mathbf{H}_{\text{tx}\text{-}\text{ris}, n_{\text{f}}}$ and $\mathbf{h}_{\text{ris}\text{-}\text{ue}, n_{\text{f}}}$ are the channels from the TX to the RIS and from the RIS to the UE, and their elements can be defined similarly to \eqref{eq-aaa} and \eqref{eq-bbb}, respectively.
$\mathbf{H}_{\text{roi}\text{-}\text{ris}, n_{\text{f}}}\in\mathbb{C}^{N_{\text{i}}\times N_{\text{s}}}$ represents the channel from the ROI to the RIS, and its $(n_{\text{i}}, n_{\text{s}})$-th element is given as
\begin{equation}\label{eq-ccc}
h_{\text{roi}\text{-}\text{ris}, n_{\text{f}}, n_{\text{i}}, n_{\text{s}}} = \frac{1}{\sqrt{4\pi} d_{n_{\text{i}}, n_{\text{s}}}}e^{-j2\pi \frac{d_{n_{\text{i}}, n_{\text{s}}}}{\lambda_{n_{\text{f}}}}},
\end{equation}
where $d_{n_{\text{i}}, n_{\text{s}}}$ is the distance between the $n_{\text{i}}$-th voxel and the $n_{\text{s}}$-th RIS element.
Unlike \eqref{eq-aaa} and \eqref{eq-bbb}, \eqref{eq-ccc} does not include transmitting or receiving gain terms because the passive RIS and ROI only scatter signals. Their scattering effects have already been captured by $G_{\text{ris}}$ and $\boldsymbol{\sigma}$, respectively.
Finally, the reciprocity of the channel between the ROI and RIS leads to $\mathbf{H}_{\text{ris}\text{-}\text{roi}, n_{\text{f}}} = \mathbf{H}_{\text{roi}\text{-}\text{ris}, n_{\text{f}}}^{\text{T}}$.

Next, we provide details of $f_{\text{phy}}$ in \eqref{eq-physical-model}.
Similar to \eqref{eq-000}, we define $\bar{\boldsymbol{\omega}}_k = G_{\text{ris}}\boldsymbol{\omega}_k$.
For the $n_{\text{f}}$-th subcarrier, we can derive ${\mathbf{H}}_{\text{sen}, n_{\text{f}}}$ according to \eqref{eq-sensing-channel}, whose components are given as
\begin{equation}
\begin{aligned}
&\mathbf{H}_{\text{tx}\text{-}\text{roi}\text{-}\text{rx}, n_{\text{f}}} \!\!\!&= &\ \mathbf{H}_{\text{tx}\text{-}\text{roi}, n_{\text{f}}}\text{diag}(\boldsymbol{\sigma})\mathbf{H}_{\text{roi}\text{-}\text{rx}, n_{\text{f}}},\\
&\mathbf{H}_{\text{tx}\text{-}\text{ris}\text{-}\text{rx}, n_{\text{f}}} \!\!\!&= &\ \mathbf{H}_{\text{tx}\text{-}\text{ris}, n_{\text{f}}}\text{diag}(\bar{\boldsymbol{\omega}}_k)\mathbf{H}_{\text{ris}\text{-}\text{rx}, n_{\text{f}}},\\
&\mathbf{H}_{\text{tx}\text{-}\text{roi}\text{-}\text{ris}\text{-}\text{rx}, n_{\text{f}}} \!\!\!&= &\ \mathbf{H}_{\text{tx}\text{-}\text{roi}, n_{\text{f}}}\text{diag}(\boldsymbol{\sigma}) \mathbf{H}_{\text{roi}\text{-}\text{ris}, n_{\text{f}}} \text{diag}(\bar{\boldsymbol{\omega}}_k) \mathbf{H}_{\text{ris}\text{-}\text{rx}, n_{\text{f}}},\\
&\mathbf{H}_{\text{tx}\text{-}\text{ris}\text{-}\text{roi}\text{-}\text{rx}, n_{\text{f}}} \!\!\!&= &\ \mathbf{H}_{\text{tx}\text{-}\text{ris}, n_{\text{f}}}\text{diag}(\bar{\boldsymbol{\omega}}_k) \mathbf{H}_{\text{ris}\text{-}\text{roi}, n_{\text{f}}} \text{diag}(\boldsymbol{\sigma}) \mathbf{H}_{\text{roi}\text{-}\text{rx}, n_{\text{f}}},
\end{aligned}
\end{equation}
where $\mathbf{H}_{\text{roi}\text{-}\text{rx}, n_{\text{f}}}\in\mathbb{C}^{N_{\text{i}}\times N_{\text{r}}}$ denotes the channel from the ROI to the RX.
The $(n_{\text{i}}, n_{\text{r}})$-th element in $\mathbf{H}_{\text{roi}\text{-}\text{rx}, n_{\text{f}}}$ is given as
\begin{equation}\label{eq-ddd}
{h}_{\text{roi}\text{-}\text{rx}, n_{\text{f}}, n_{\text{i}}, n_{\text{r}}} = \lambda_0\sqrt{\frac{G_{\text{r}}}{4\pi}} \times \frac{1}{\sqrt{4\pi} d_{n_{\text{i}}, n_{\text{r}}}}e^{-j2\pi \frac{d_{n_{\text{i}}, n_{\text{r}}}}{\lambda_{n_{\text{f}}}}},
\end{equation}
where $G_{\text{r}}$ is the antenna gain of the RX, and $d_{n_{\text{i}}, n_{\text{r}}}$ is the distance from the $n_{\text{i}}$-th voxel to the $n_{\text{r}}$-th RX antenna.
$\mathbf{H}_{\text{ris}\text{-}\text{rx}, n_{\text{f}}}$ represents the channel from the RIS to the RX, and its definition is similar to \eqref{eq-ddd}.
Finally, we obtain $\boldsymbol{\mathcal{H}}_{k}$ by stacking ${\mathbf{H}}_{\text{sen}, n_{\text{f}}}$ for $n_{\text{f}} = 1, \ldots, N_{\text{f}}$.

\section{}\label{appendix-optimization}

In this section, we formulate the RIS phase optimization problem that maximizes the SE of the central subcarrier frequency, denoted as the $n_0$-th subcarrier.
First, we derive
\begin{equation*}
\begin{aligned}
&\mathbf{h}_{\text{a}, n_0} = \mathbf{h}_{\text{tx}\text{-}\text{ue}, n_0} + \mathbf{h}_{\text{tx}\text{-}\text{roi}\text{-}\text{ue}, n_0}, \\
&\mathbf{h}_{\text{b}, n_0} = \mathbf{h}_{\text{tx}\text{-}\text{ris}\text{-}\text{ue}, n_0} + \mathbf{h}_{\text{tx}\text{-}\text{ris}\text{-}\text{roi}\text{-}\text{ue}, n_0} + \mathbf{h}_{\text{tx}\text{-}\text{roi}\text{-}\text{ris}\text{-}\text{ue}, n_0}.
\end{aligned}
\end{equation*}
Then, we have $\mathbf{h}_{\text{b}, n_0} = \mathbf{H}_{\text{b}, n_0}\boldsymbol{\omega}$, where
\begin{equation*}
\begin{aligned}
& \mathbf{h}_{\text{c}, n_0} \!\!\!&= &\ \mathbf{h}_{\text{ue}\text{-}\text{ris}, n_0} + \mathbf{H}_{\text{roi}\text{-}\text{ris}, n_0} \text{diag}(\boldsymbol{\sigma}) \mathbf{h}_{\text{ue}\text{-}\text{roi}, n_0},\\
& \mathbf{H}_{\text{a}, n_0} \!\!\!&= &\ \mathbf{H}_{\text{roi}\text{-}\text{tx}, n_0} \text{diag}(\boldsymbol{\sigma}) \mathbf{H}_{\text{ris}\text{-}\text{roi}, n_0},\\
& \mathbf{H}_{\text{b}, n_0} \!\!\!&= &\ \mathbf{H}_{\text{ris}\text{-}\text{tx}, n_0} \text{diag}(\mathbf{h}_{\text{c}, n_0}) + \mathbf{H}_{\text{a}, n_0} \text{diag}(\mathbf{h}_{\text{ue}\text{-}\text{ris}, n_0}).
\end{aligned}
\end{equation*}
According to \eqref{eq-se1}, the RIS phase optimization problem to maximize the SE can be formulated as
\begin{equation*}
\begin{aligned}
\text {(P1):}\ \  & \max \limits_{\boldsymbol{\omega}} && \left\|\mathbf{h}_{\text{com}, n_0}\right\|^{2} = \left\|\mathbf{H}_{\text{b}, n_0}\boldsymbol{\omega} + \mathbf{h}_{\text{a}, n_0}\right\|^{2}, \\
& \text { s.t. } && 0 \leq \omega_{n_{\text{s}}} \leq 2 \pi, \ \ \forall n_{\text{s}}=1, \ldots, N_\text{s}. \\
\end{aligned}
\end{equation*}
Consequently, the problem (P1) can be solved by exploiting the RIS phase optimization algorithm proposed in \cite{wu2018intelligent}, deriving the sub-optimal RIS configurations for communication.

\end{appendices}

\bibliographystyle{IEEEtran}
\bibliography{trans_ref}{}

\vfill

\end{document}